\documentclass[12pt,a4paper]{article}
\usepackage{amsmath}
\usepackage{amssymb}
\usepackage{graphicx,epsfig}
\usepackage[mathscr]{eucal}
\usepackage{color}
\usepackage[numbers,sort&compress]{natbib}
\usepackage[english]{babel}

\usepackage{tabularx}

\usepackage{setspace}
\usepackage{xspace}

\usepackage{multirow}

\def\ltap{\raisebox{-.4ex}{\rlap{$\,\sim\,$}} \raisebox{.4ex}{$\,<\,$}}

\newcommand\as{\alpha_{\mathrm{S}}}

\def\to{\rightarrow}
\def\nn{\nonumber}

\def\rcut{r_{\rm cut}}

\def\ttb{\ensuremath{t {\bar t}}\xspace}
\def\bbb{\ensuremath{b {\bar b}}\xspace}

\newcommand\Matrix{{\sc Matrix}}

\newcommand{\rescaletwoplots}{0.48\textwidth}
\newcommand{\hspacebetweentwoplots}{\hfill}

\newcommand{\rescalethreeplots}{0.32\textwidth}
\newcommand{\hspacebetweenthreeplots}{\hfill}

\newcommand{\spaceabovefigurecaption}{\vspace*{-2ex}}

\usepackage{array}
\newcolumntype{L}[1]{>{\raggedright\let\newline\\\arraybackslash\hspace{0pt}}m{#1}}
\newcolumntype{C}[1]{>{\centering\let\newline\\\arraybackslash\hspace{0pt}}m{#1}}
\newcolumntype{R}[1]{>{\raggedleft\let\newline\\\arraybackslash\hspace{0pt}}m{#1}}

\usepackage{pstricks}
\usepackage[colorlinks=true,allcolors={blue!70!black}]{hyperref}
\usepackage{cleveref}

\usepackage{footnotebackref}

\usepackage[section]{placeins}

\usepackage{etoolbox}
\def\appendixname{Appendix}% don't add the space here
\appto\appendix{%
  \addtocontents{toc}{\patch@l@section}% patch \l@section
  \appto\appendixname{ }% here add a trailing space
}
\protected\def\patch@l@section{%
  \patchcmd{\l@section}{1.5em}{\widthof{\appendixname\space}+2.5em}{}{}%
}
\begin{document}
\begin{titlepage}
\begin{flushright}
ZU-TH 36/20\\
TIF-UNIMI-2020-27\\
MPP-2020-186
\end{flushright}

\renewcommand{\thefootnote}{\fnsymbol{footnote}}
\vspace*{0.5cm}

\begin{center}
  {\Large \bf Bottom-quark production at hadron colliders: \\[0.4cm]
  fully differential predictions in NNLO QCD}
\end{center}

\par \vspace{2mm}
\begin{center}
  {\bf Stefano Catani${}^{(a)}$, Simone Devoto${}^{(b,c)}$, Massimiliano Grazzini${}^{(b)}$,\\[0.2cm]
    Stefan Kallweit${}^{(d)}$}
and
{\bf Javier Mazzitelli${}^{(e)}$}

\vspace{5mm}

${}^{(a)}$INFN, Sezione di Firenze and
Dipartimento di Fisica e Astronomia,\\
Universit\`a di Firenze,
50019 Sesto Fiorentino, Firenze, Italy\\[0.25cm]

${}^{(b)}$Physik Institut, Universit\"at Z\"urich, 8057 Z\"urich, Switzerland\\[0.25cm]

${}^{(c)}$Dipartimento di Fisica, Universit\`{a} degli Studi di Milano, 20133 Milano, Italy\\[0.25cm]

$^{(d)}$Dipartimento di Fisica, Universit\`{a} degli Studi di Milano-Bicocca and\\
INFN, Sezione di Milano-Bicocca,
20126 Milano, Italy\\[0.25cm]

${}^{(e)}$Max-Planck-Institut f\"ur Physik, F\"ohringer Ring 6, 80805 M\"unchen, Germany

\vspace{5mm}

\end{center}

\par \vspace{2mm}
\begin{center} {\large \bf Abstract}

\end{center}
\begin{quote}
\pretolerance 10000

We report on the first fully differential calculation of the next-to-next-to-leading-order (NNLO) QCD radiative corrections to the production of bottom-quark pairs at hadron colliders.
The calculation is performed by using the $q_T$ subtraction formalism to handle and cancel infrared singularities in real and virtual contributions.
The computation is implemented in the \Matrix\ framework, thereby allowing us to efficiently compute arbitrary infrared-safe observables in the four-flavour scheme.
We present selected predictions for bottom-quark production at the Tevatron and at the LHC at different collider energies, and we perform some comparisons with available experimental results.
We find that the NNLO corrections are sizeable, typically of the order of $25$--$35\%$, and they lead to a significant reduction of the perturbative uncertainties.
Therefore, their inclusion is crucial for an accurate theoretical description of this process.

\end{quote}

\vspace*{\fill}
\begin{flushleft}
October 2020
\end{flushleft}
\end{titlepage}

%=============================================================================================
\section{Introduction}
\label{sec:intro}
The production of bottom quarks has been extensively studied at hadron colliders. Early measurements were already carried out by the UA1 collaboration at the CERN S$p{\bar p}$S~\cite{Albajar:1986iu,Albajar:1988th}
and, subsequently,
by the CDF~\cite{Abe:1995dv,Acosta:2001rz,Acosta:2004yw,Abulencia:2006ps} and D0~\cite{Abachi:1994kj,Abbott:1999se} collaborations at the Fermilab Tevatron. The most recent measurements were performed by the ALICE~\cite{Abelev:2014hla,Abelev:2012sca}, ATLAS~\cite{Aad:2012jga, ATLAS:2013cia}, CMS~\cite{Khachatryan:2011mk,Chatrchyan:2011pw,Khachatryan:2016csy,Chatrchyan:2012hw}, and LHCb~\cite{Aaij:2010gn,Aaij:2012jd,Aaij:2013noa,Aaij:2016avz} collaborations at the CERN LHC in $pp$ collisions at centre-of-mass energies $\sqrt{s}=2.76,7,8$ and 13 TeV.

At the theoretical level, heavy-quark production at hadron colliders is one of the most classic tests of perturbative QCD.
The cross section to produce a pair of heavy quarks with mass $m_Q$ is computable as a power series expansion in the QCD coupling $\as(\mu_R)$,
where the renormalisation scale $\mu_R$ has to be chosen of the order of $m_Q$.
In the case of the bottom ($b$) quark the relatively low mass, $m_b\sim 4-5$ GeV, leads to a slow convergence of the perturbative expansion and, therefore, to large theoretical uncertainties.

Theoretical predictions for $b{\bar b}$ production at next-to-leading order (NLO) in QCD have been available
for a long time~\cite{Nason:1987xz,Nason:1989zy,Beenakker:1988bq},
including calculations~\cite{Mangano:1991jk} of $b{\bar b}$ correlations and generic
infrared safe (IR) observables.
NLO studies based on different schemes for the renormalisation of the
bottom-quark mass are presented in Ref.~\cite{Garzelli:2020fmd}.
At high transverse momenta $p_T$ of the bottom quark, large logarithmic terms of the form $\ln (p_T/m_b)$
need to be resummed to all perturbative orders~\cite{Cacciari:1993mq}.
In the case of single-particle ($b$ quark or antiquark) inclusive cross sections
the resummation can be performed by introducing the perturbative fragmentation function~\cite{Mele:1990cw}
of the bottom quark (which can also be supplemented with all-order soft-gluon effects~\cite{Cacciari:2001cw}).
Predictions obtained by matching such resummed computations to the NLO calculation (the so called ``FONLL'' prediction)~\cite{Cacciari:1998it, Cacciari:2001td,
Cacciari:2002pa, Cacciari:2012ny}
have become the standard reference for the comparison with experimental data. Perturbative predictions for the inclusive production of bare bottom quarks can then be folded with non-perturbative functions
\cite{Kartvelishvili:1977pi, Peterson:1982ak}
describing the fragmentation into the triggered $b$-hadrons.
The parameters that control such fragmentation functions are typically extracted from LEP data (see, e.g., Ref.~\cite{Cacciari:2005uk}).
The variable-flavour-number scheme~\cite{Kniehl:2004fy, Kniehl:2005mk} is another
procedure that is used to combine the NLO calculation with high-$p_T$ resummation
effects for single-inclusive $b$-hadron production, and ensuing data--theory comparisons have been presented in the literature (see, e.g.,
Refs.~\cite{Kramer:2018vde,Benzke:2019usl}).
Phenomenological studies on the impact of bottom
production measurements on parton distribution functions have been presented in Refs.~\cite{Zenaiev:2015rfa,Cacciari:2015fta,Zenaiev:2019ktw}.

At the next-to-next-to-leading order (NNLO) in QCD, theoretical predictions for $b{\bar b}$ production are available only for the total cross section.
Indeed, the $b\bar b$ results can be directly derived by exploiting the corresponding NNLO theoretical calculation~\cite{Baernreuther:2012ws,Czakon:2012zr,Czakon:2012pz,Czakon:2013goa} of the total cross section for top-quark pair production.
NNLO values of the $b{\bar b}$ total cross section at several collider energies are presented in Refs.~\cite{Mangano:2016jyj,dEnterria:2016ids}.
The calculation of the $b{\bar b}$ total cross section at NNLO is implemented in the numerical program {\sc Hathor}~\cite{Langenfeld:2009wd,Aliev:2010zk}.

In this paper we report on the first fully differential NNLO QCD calculation for $b{\bar b}$ production at hadron colliders, and we also present comparisons with some inclusive $b$-hadron data obtained at the Tevatron and at the LHC.
Our computation, which follows the analogous calculation carried out for top-quark pair production~\cite{Catani:2019iny,Catani:2019hip,Catani:2020tko}, is implemented within the {\sc Matrix} framework~\cite{Grazzini:2017mhc} and allows us to evaluate arbitrary IR safe observables for the production of on-shell $b$ and ${\bar b}$ quarks at hadron colliders.

The NNLO calculation requires tree-level, one-loop and two-loop contributions.
We compute the tree-level and one-loop scattering amplitudes by using {\sc OpenLoops}~\cite{Cascioli:2011va,Buccioni:2017yxi,Buccioni:2019sur}.
We use the numerical result of Ref.~\cite{Baernreuther:2013caa} to evaluate the two-loop amplitudes.
The IR divergences that appear at intermediate stages of the computation are handled and cancelled, analogously to Refs.~\cite{Catani:2019iny,Catani:2019hip}, by using the $q_T$-subtraction formalism~\cite{Catani:2007vq}, which was properly extended to deal with heavy-quark production in Refs.~\cite{Catani:2014qha,Bonciani:2015sha}.

The paper is organised as follows. In Section~\ref{sec:matrix} we briefly review the {\sc Matrix} framework applied to heavy-quark production.
In Section~\ref{sec:resu} we present our numerical results: we first discuss the total cross section in Section~\ref{sec:resu:tota}, and then we report results for differential distributions
at the Tevatron in Section~\ref{sec:resu:tev} and at the LHC in Section~\ref{sec:resu:lhc}. Our results are summarised in Section~\ref{sec:summa}.
We devote \ref{app:eta-y} to a detailed discussion of the shape differences between
rapidity and pseudorapidity distributions.
In \ref{app:FONLL} we present a comparison of our NNLO results with FONLL predictions for the transverse-momentum and (pseudo)rapidity distributions of the bottom quark.
%=============================================================================================

%=============================================================================================
\section[{\sc Matrix} framework for heavy-quark production]{M{\normalsize ATRIX} framework for heavy quark production}
\label{sec:matrix}
The results presented in this work are obtained by using the $q_T$-subtraction formalism~\cite{Catani:2007vq} to handle and cancel IR singularities.
Specifically, we use the implementation of the formalism within the computational framework {\sc Matrix}.
In its public version, {\sc Matrix} permits the evaluation of differential distributions at NNLO in QCD for a wide class of processes in which the triggered final state is formed by colourless particles.

Recently, the computation of the last ingredients (namely NNLO soft gluon contributions) needed to extend $q_T$ subtraction to heavy-quark production was completed by some of us~\cite{inprep}.
An independent computation of these contributions is presented in Ref.~\cite{Angeles-Martinez:2018mqh} for the case of top-quark pair production.
With the results of Ref.~\cite{inprep} we were able to carry out a new calculation of top-quark pair production at NNLO in QCD~\cite{Catani:2019iny},
completing a previous work~\cite{Bonciani:2015sha} that was limited to the flavour off-diagonal production channels.
Their integration in the {\sc Matrix} framework allowed us to perform an efficient evaluation of single- and multi-differential distributions for stable top quarks~\cite{Catani:2019hip}.
For the present work we have generalised this implementation to arbitrary heavy-quark mass values and light-flavour numbers.
We apply a scheme with $n_f=4$ light-quark flavours and a massive bottom quark, while the top quark is decoupled from the process, to
obtain differential NNLO results for bottom-quark pair production.

The NNLO differential cross section for bottom-pair production, $d{\sigma}^{b{\bar b}}_{\rm NNLO}$, is obtained within the $q_T$-subtraction method according to the following main formula,
\begin{equation}
\label{eq:main}
d{\sigma}^{b{\bar b}}_{\rm NNLO}={\cal H}^{b{\bar b}}_{\rm NNLO}\otimes d{\sigma}^{b{\bar b}}_{\rm LO}
+\left[ d{\sigma}^{b{\bar b}+\rm{jet}}_{\rm NLO}-
d{\sigma}^{b{\bar b}, \, {\rm CT}}_{\rm NNLO}\right],
\end{equation}
where $d{\sigma}^{b{\bar b}+\rm{jet}}_{\rm NLO}$ represents the $\bbb$+jet cross section at NLO accuracy, which we
evaluate by using the dipole subtraction method~\cite{Catani:1996jh,Catani:1996vz,Catani:2002hc}.
As is customary in the {\sc Matrix} framework, the NLO cross section $d\sigma^{b{\bar b}}_{\rm NLO}$ is also computed by using dipole subtraction, while $q_T$ subtraction is actually used only to evaluate
the NNLO correction $d\sigma^{b{\bar b}}_{\rm NNLO}-d\sigma^{b{\bar b}}_{\rm NLO}$.

The expression in Eq.~(\ref{eq:main}) is completely analogous to the \ttb case~\cite{Catani:2019hip}.
We remind the reader that the term in the square bracket of Eq.~(\ref{eq:main}) is formally finite in the limit $q_T\to 0$ ($q_T$ is the transverse momentum of the $b\bar b$ pair),
but each of the two contributions in the square bracket is individually divergent.
Therefore a technical cut is introduced in the dimensionless quantity $r=q_T/m_{\bbb}$ ($m_{\bbb}$ is the invariant mass of the $b{\bar b}$ pair).
The $r_\text{cut}\to 0$ extrapolation is performed following the procedure of Ref.~\cite{Grazzini:2017mhc}, and it is also applied in the computation of differential distributions on a bin-by-bin basis.

The core of {\sc Matrix} is the Monte Carlo program {\sc Munich}\footnote{{\sc Munich}, which is the abbreviation of “MUlti-chaNnel Integrator at Swiss (CH) precision”, is an automated parton-level NLO generator by S. Kallweit.}, which contains a fully automated implementation of the dipole subtraction method for massless~\cite{Catani:1996jh,Catani:1996vz} and massive~\cite{Catani:2002hc} partons and a general implementation of an efficient phase space integration.
The required spin- and colour-correlated tree-level and one-loop amplitudes are obtained by using {\sc OpenLoops}~\cite{Cascioli:2011va, Buccioni:2017yxi,Buccioni:2019sur}, with the exception of the four-parton tree-level colour correlators for which we rely on an analytic implementation.
The two-loop amplitudes are obtained via an interpolation routine, based on the numerical results presented in Refs.~\cite{Czakon:2008zk, Baernreuther:2013caa}.
%=============================================================================================

%=============================================================================================
\section{Results}
\label{sec:resu}
In the following we present our results for total cross sections and differential distributions for $b{\bar b}$ production at the Tevatron ($\sqrt{s}=1.96$ TeV) and at the LHC ($\sqrt{s}=7$ and $13$ TeV).
We use the NNPDF31 parton distribution functions~(PDFs)~\cite{Ball:2017nwa} with $n_f=4$ massless-quark flavours and the value of the QCD coupling $\as(m_Z)=0.118$.
The pole mass of the bottom quark is fixed to $m_b=4.92$~GeV, consistently with the value that is used in the chosen PDF set.
Predictions at the $n$-th perturbative order are obtained by using the corresponding N$^n$LO PDF set and the evolution of $\as$ at $(n+1)$-loops.
Since the NNPDF31 set is not available at LO with $n_f=4$, we instead use the corresponding NNPDF30 set~\cite{Ball:2014uwa} for our LO predictions.
Perturbative uncertainties are estimated with the customary 7-point variation of the renormalisation ($\mu_R$) and factorisation ($\mu_F$) scales by a
factor of two around a common central value $\mu_0$,
with the constraint $0.5 \leq \mu_R/\mu_F \leq 2$.
The value of $\mu_0$ is chosen of the order of the characteristic hard-scattering scale of the process.
The total cross section is controlled by scales of the order of $m_b$, while each differential distribution is characterised by a different hard-scattering scale that has to be specified accordingly.
The fully differential nature of our calculation also allows us to use dynamic scales.

We point out that starting from NNLO the inclusive production of a $b{\bar b}$ pair receives contributions from tree-level diagrams with an additional $b\bar b$ pair in the final state.
Since the bottom-quark mass is kept non vanishing, these four-bottom contributions are separately finite and may be included or not in the perturbative calculation,
according to the actual (theoretical and experimental) definition of the inclusive cross section.
We find that these contributions are generally rather small. They are completely negligible at the Tevatron while
at the LHC they typically contribute at the per mille level, reaching about $0.5\%$ at large transverse momenta of the bottom quarks.
Owing to their small size, these contributions have no visible quantitative effects on the results that we are going to present.

Before presenting our numerical predictions, we discuss the impact of the two-loop virtual corrections on our NNLO results. The IR finite part of the two-loop contribution
is obtained by using the numerical results of Ref.~\cite{Baernreuther:2013caa}, which are provided through a $80\times 21$ dimensional grid in the Born level kinematical variables $\beta$ and $\cos\theta$ ($\beta$ and $\theta$ are the heavy-quark velocity and scattering angle in the partonic centre-of-mass frame).
The grid is then interpolated by using splines, and the ensuing result is supplemented with the analytical results at small and large $\beta$~\cite{Beneke:2009ye,Baernreuther:2013caa}.
The grid of Ref.~\cite{Baernreuther:2013caa} has, to date, been used only for top-quark production.
Since the bottom-quark mass is significantly smaller than the top-quark mass, one may wonder whether the small-angle (collinear) region is sampled sufficiently well in our calculation.
We have studied the effect of the IR finite part of the two-loop contribution on our calculations of the total cross section and several differential distributions for $b{\bar b}$ production.
Moreover, to quantify the impact of having a discrete grid,
we have repeated our calculations by using only half of the available grid points.
We find that the two-loop virtual contribution~\cite{Baernreuther:2013caa} is below $1\%$ in all cases. The differences obtained by reducing the number of points in the grid by a factor of two are
typically at the per mille level for all the distributions we have considered.
We conclude that we can safely use the results of Ref.~\cite{Baernreuther:2013caa} to carry out our fully differential calculation of \bbb production.

\subsection{Total cross section}
\label{sec:resu:tota}

We start the presentation of our results by considering the total cross section.
The total cross section for the production of a pair of bottom quarks is controlled by scales of the order of the bottom mass $m_b$.
Accordingly, we will consider the two central scales $\mu_0=m_b$ and $\mu_0=2m_b$.

\begin{figure}[t]
\centering
\includegraphics[width=\rescaletwoplots]{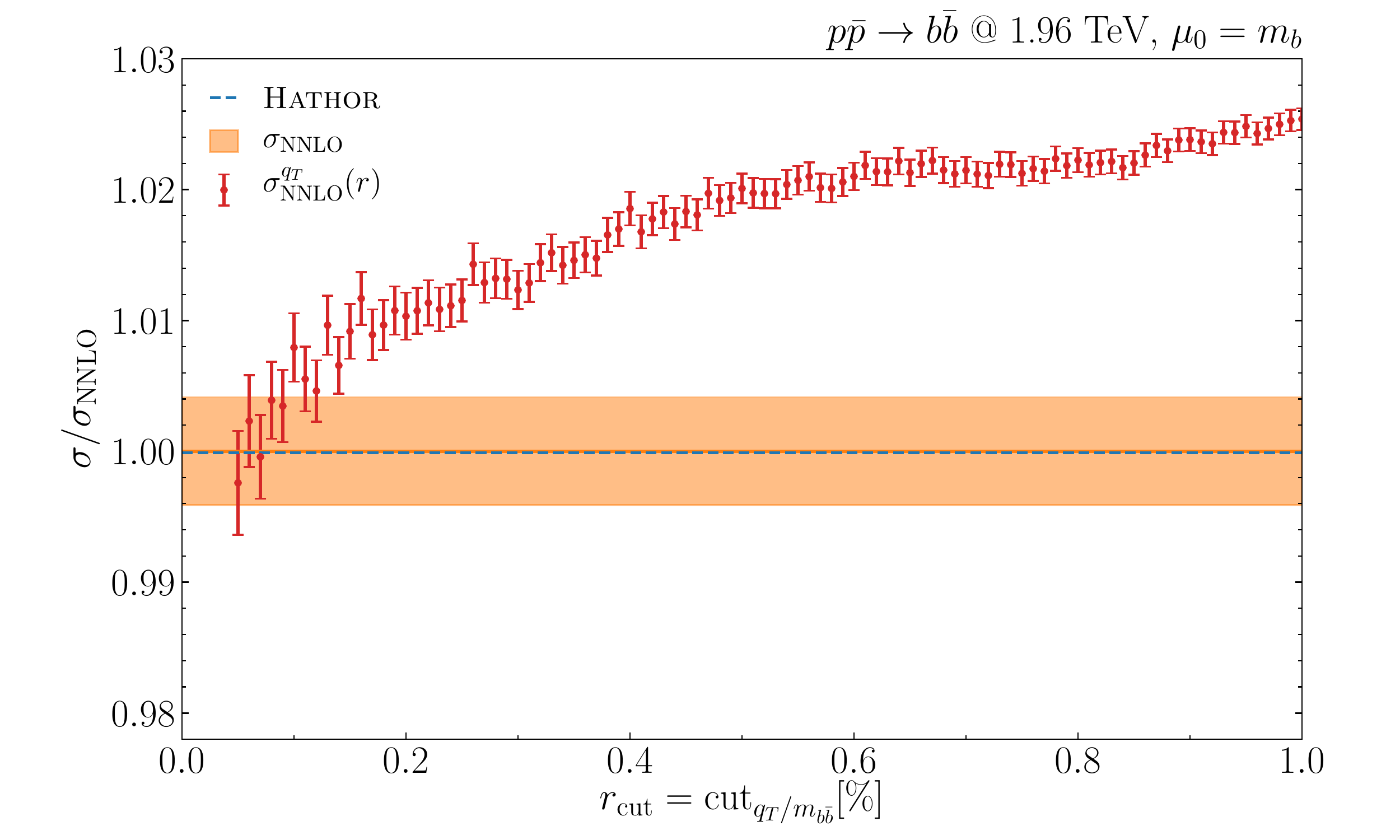}\hspacebetweentwoplots
\includegraphics[width=\rescaletwoplots]{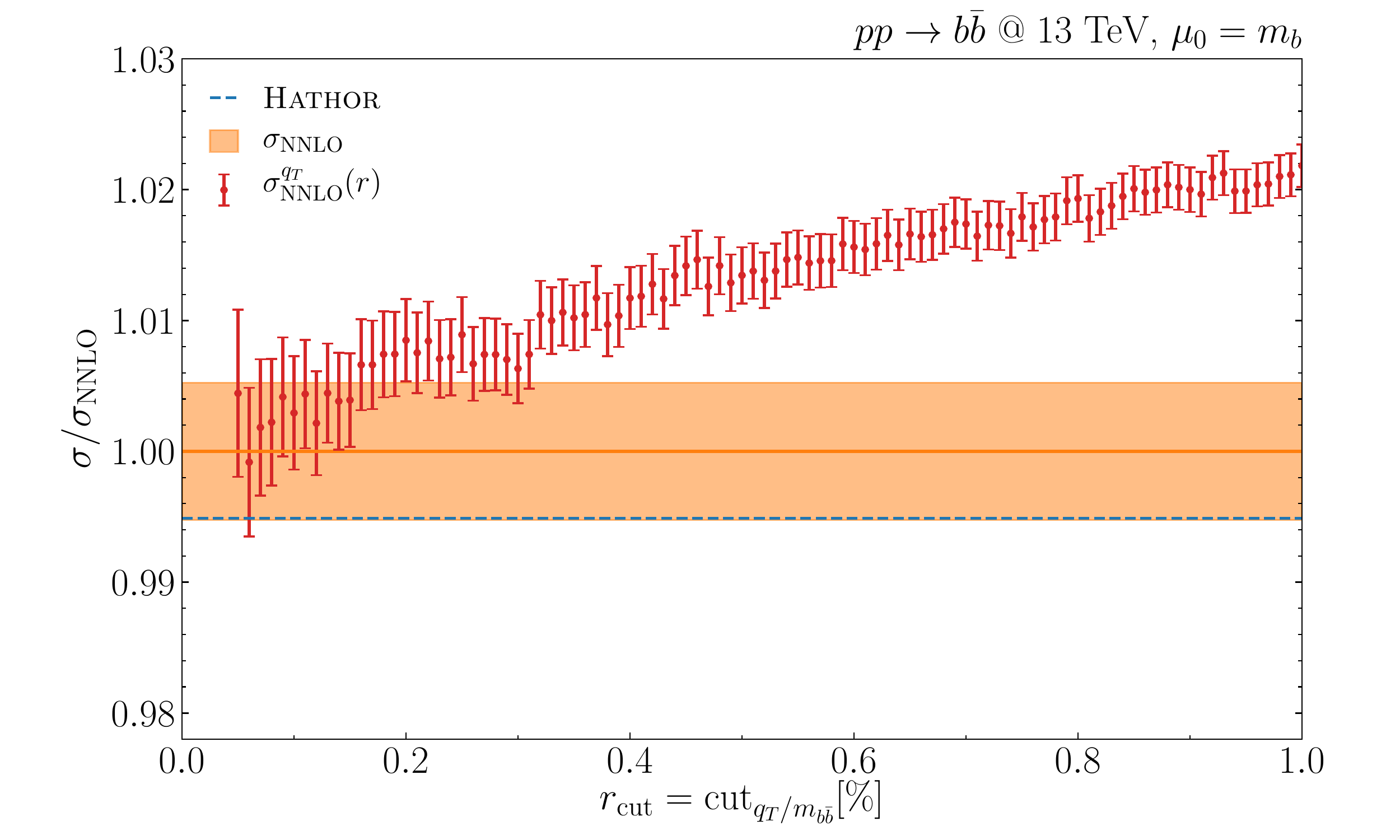}
\spaceabovefigurecaption
\caption{The dependence on $r_{\rm cut}$ of the total cross section for $b\bar{b}$ production at different hadron colliders.}
\label{fig:r_cut}
\end{figure}

As discussed in Sec.~\ref{sec:matrix}, our NNLO results are obtained through an $\rcut\to 0$ extrapolation procedure.
The $\rcut$ dependence at the Tevatron with $\sqrt{s}=1.96$ TeV and at the LHC with $\sqrt{s}=13$ TeV is shown in Fig.~\ref{fig:r_cut} in the case $\mu_0=m_b$.
The ensuing NNLO cross section with its extrapolation uncertainty is compared with the corresponding result obtained with the numerical program {\sc Hathor}~\cite{Aliev:2010zk},
and the results of the two NNLO calculations are in good agreement. Similar results are obtained for all scale combinations
of the applied 7-point variation, and also separated into partonic channels as in Ref.~\cite{Catani:2019iny}.
Fig.~\ref{fig:r_cut} shows that the extrapolation uncertainties are larger than the corresponding uncertainties in the case of top-pair production~\cite{Catani:2019iny}, but, remarkably, still at the level of about $0.5\%$.
This larger uncertainty in the NNLO result is simply due to the larger relative size of the ${\cal O}(\as^4)$ contribution when considering a lower quark mass. Nevertheless, this level of uncertainty is perfectly acceptable as it is well below other sources of theoretical (and experimental) uncertainties affecting bottom-quark hadroproduction.

The NNLO results at the Tevatron and the LHC obtained with $\mu_0=m_b$ and $\mu_0=2m_b$ together with their scale uncertainties are reported in Table~\ref{table:total_vs_hathor}, compared to those obtained using {\sc Hathor}~\cite{Aliev:2010zk}. For both central-scale choices the agreement is excellent, including scale uncertainties.
We point out that the two computations are performed by using fully independent methods.

\begin{table}[tb]
\begin{center}
\renewcommand{\arraystretch}{1.6}
\setlength{\tabcolsep}{1em}
\begin{tabular}{|l|c|c|c|}
\hline
\multicolumn{1}{|c|}{$\sigma_{\rm NNLO}\;[\mu\text{b}]$}&\multicolumn{1}{c|}{$p \bar{p}$ @ $1.96$ TeV} & \multicolumn{1}{c|}{$pp$ @ $7$ TeV} & \multicolumn{1}{c|}{$pp$ @ $13$ TeV}\\ \hline
\multicolumn{4}{|c|}{\textbf{$\mu_0=m_b$}}  \\ \hline
{\sc Matrix} & $75.4(3)\phantom{.}^{+22\%}_{-21\%}$ & $288(2)\phantom{.}^{+30\%}_{-24\%}$ & $508(3)\phantom{.}^{+32\%}_{-25\%}$\\ 
{\sc Hathor} & $75.45\phantom{().}^{+22\%}_{-21\%}$ & $284.3\phantom{()}^{+30\%}_{-24\%}$  &$505.5\phantom{()}^{+32 \%}_{-25\%}$\\\hline
\multicolumn{4}{|c|}{\textbf{$\mu_0=2m_b$}}  \\ \hline
{\sc Matrix} & $66.7(2)\phantom{.}^{+ 21\%}_{-18 \%}$ & $258(1)\phantom{.}^{+ 20\%}_{-18 \%}$ & $458(2)\phantom{.}^{+ 20\%}_{-18 \%}$\\ 
{\sc Hathor} & $66.70\phantom{().}^{+21\%}_{-18\%}$ & $256.1\phantom{()}^{+19\%}_{-18\%}$ & $455.8\phantom{()}^{+19\%}_{-18\%}$\\\hline
\end{tabular}
\end{center}
\caption{
The NNLO total cross section for $b\bar{b}$ production at the Tevatron and at the LHC: comparison of our results with those obtained by using {\sc Hathor}.
The quoted uncertainties are obtained through scale variations as described in the text. Numerical errors on the last
digit are stated in brackets (and include the uncertainties due to the $\rcut\to 0$ extrapolation). 
}
\label{table:total_vs_hathor}
\end{table}

In Table~\ref{table:totalXS} we report the LO, NLO and NNLO results for $\mu_0=m_b$ and $\mu_0=2m_b$.
As in Table~\ref{table:total_vs_hathor}, the cross sections are presented with their perturbative uncertainties estimated through scale variations.

\begin{table}[tb]
\begin{center}
\renewcommand{\arraystretch}{1.6}
\setlength{\tabcolsep}{1em}
\begin{tabular}{|l|c|c|c|}
\hline
\multicolumn{1}{|c|}{$\sigma\;[\mu\text{b}]$}&\multicolumn{1}{c|}{$p \bar{p}$ @ $1.96$ TeV} & \multicolumn{1}{c|}{$pp$ @ $7$ TeV} & \multicolumn{1}{c|}{$pp$ @ $13$ TeV} \\ \hline
\multicolumn{4}{|c|}{\textbf{$\mu_0=m_b$}} \\ \hline
LO   & $34.66\phantom{().}^{+51\%}_{-32\%}$ & $138.7\phantom{()}^{+51\%}_{-46\%}$ & $249.0\phantom{()}^{+59\%}_{-51\%}$ \\ 
NLO  & $60.23\phantom{().}^{+54\%}_{-28\%}$ & $219.8\phantom{()}^{+61\%}_{-39\%}$ & $378.6\phantom{()}^{+65\%}_{-45\%}$ \\ 
NNLO & $75.4(3)\phantom{.}^{+22\%}_{-21\%}$ & $288(2)\phantom{.}^{+30\%}_{-24\%}$ & $508(3)\phantom{.}^{+32\%}_{-25\%}$ \\ 
%  {\sc Hathor} & $75.45^{+22\%}_{-21\%}$ & $284.3^{+30\%}_{-24\%}$  &$505.5^{+32 \%}_{-25\%}$\\
\hline
\multicolumn{4}{|c|}{\textbf{$\mu_0=2m_b$}} \\ \hline
LO   & $30.94\phantom{().}^{+41\%}_{-25\%}$ & $145.8\phantom{()}^{+41\%}_{-32\%}$ & $281.9\phantom{()}^{+41\%}_{-37\%}$ \\ 
NLO  & $51.16\phantom{().}^{+33\%}_{-23\%}$ & $203.3\phantom{()}^{+36\%}_{-26\%}$ & $362.9\phantom{()}^{+34\%}_{-28\%}$ \\ 
NNLO & $66.7(2)\phantom{.}^{+21\%}_{-18\%}$ & $258(1)\phantom{.}^{+20\%}_{-18\%}$ & $458(2)\phantom{.}^{+20\%}_{-18\%}$ \\ 
%  {\sc Hathor} &$66.70^{+21\%}_{-18\%}$  &$256.1^{+19\%}_{-18\%}$   &$455.8^{+19\%}_{-18\%}$\\
\hline
\end{tabular}
\end{center}
\vspace*{-0.4cm}
\caption{
Total cross section for $b\bar{b}$ production at LO, NLO and NNLO.
The quoted uncertainties are obtained through scale variations as described in the text. 
At NNLO the numerical errors on the last digits are stated in brackets, as in Table~\ref{table:total_vs_hathor}.
}
\label{table:totalXS}
\end{table}

By inspecting the results presented in Table~\ref{table:totalXS}, we immediately see that QCD corrections are very large.
In order to quantify their impact, we introduce $K$-factors, $K_\text{(N)NLO} = \sigma_\text{(N)NLO}/\sigma_\text{(N)LO}$, defined as the ratios of the cross section predictions at two subsequent orders.
The NLO $K$-factor $K_\text{NLO}$ tends to increase as the collider energy decreases. Specifically, the value of $K_\text{NLO}$
ranges between 1.29 (LHC, $\sqrt s =13$ TeV) and 1.74 (Tevatron) for the different energies and central scales under consideration.
The NNLO corrections are still sizeable, but weakly depend on the collider energy, with values of $K_\text{NNLO}$ ranging from 1.25 to 1.34.
These large QCD corrections, considerably larger for instance than the ones observed in top-quark production, are associated to the relatively low energy scales involved, which lead to large values of the strong coupling.
The poor perturbative behaviour is also reflected by the large scale uncertainties that are observed in Table~\ref{table:totalXS}.
It is important to remark, however, that in all cases the inclusion of the NNLO contribution allows us to strongly reduce the theoretical uncertainties.
In addition, $K_\text{NNLO}$ is always smaller than $K_\text{NLO}$, providing a sign of (slow) convergence of the perturbative series
We also note that the values of $\sigma_{\text{NNLO}}$ and $\sigma_{\text{NLO}}$ are consistent within their scale uncertainites for each energy and choice of $\mu_0$ (the values of $\sigma_{\text{NLO}}$ and $\sigma_{\text{LO}}$ are consistent as well).

We now turn to the discussion of the scale dependence of our results.
Due to the overall proportionality of the cross section to the factor $\as^2(\mu_R)$ at LO, the cross section typically decreases as $\mu_R$ increases.
On the contrary, the cross section generally increases as $\mu_F$ increases.
This is due to the fact that bottom-quark hadroproduction is sensitive to relatively low momentum fractions of the colliding partons, and in this kinematical region PDF scaling violations are typically positive.
When performing scale variations, the dominant effect is given by variations of the renormalization scale $\mu_R$.

Comparing the predictions corresponding to the two different scale choices, we observe that the results obtained with $\mu_0 = 2m_b$ present smaller scale uncertainties. Such difference is evident at LO and NLO, and less noticeable at NNLO, especially at the Tevatron.
The choice $\mu_0=2m_b$ leads to slightly smaller NNLO $K$-factors at the LHC ($K_\text{NNLO}=1.27(1.26)$ to be compared with $K_\text{NNLO}=1.31(1.34)$ at $\sqrt{s}=7(13)$ TeV for $\mu_0=m_b$), and
to larger $K$-factors at the Tevatron ($K_\text{NNLO}=1.30$ to be compared with $K_\text{NNLO}=1.25$ for $\mu_0=m_b$).
Despite the differences between the results obtained with these two central scales, both choices provide predictions that are fully compatible within scale uncertainties, and are equally acceptable on general grounds.
The use of $\mu_0 = m_b$ as central scale choice leads to larger uncertainties due to the lower values of $\mu_R$ involved, but we do not observe a particularly worrisome perturbative behaviour for this scale choice, in spite of the low value of $m_b$.

Besides the uncertainties related to missing higher-orders in the perturbative expansion, additional uncertainties on the bottom-quark cross section arise from the errors in the determination of the pole mass $m_b$, the parton distributions, and the strong coupling $\as(m_Z)$.
In Table \ref{table:totalXSwithunc} we report our result for the  NNLO cross section computed with $\mu_0=m_b$ including these additional uncertainties.
As for the bottom mass, we follow Ref.~\cite{deFlorian:2016spz}, and we vary $m_b$ between $m_b=4.79$ GeV and $m_b=5.05$ GeV, corresponding to $m_b=4.92\pm 0.13$ GeV.
The effect of changing the bottom mass
is below $10\%$ and slightly decreases with increasing collider energy. The PDF uncertainties are much smaller and increase with the collider energy.
This is not unexpected: indeed,
$b{\bar b}$ production at the Tevatron and the LHC is mainly sensitive to PDFs with momentum fractions $x\sim 2m_b/\sqrt{s}$ in the range ${\cal O}(5\cdot 10^{-3})-{\cal O}(10^{-3})$. In this region the uncertainty in the gluon distribution increases as $x$ decreases.
As for the QCD coupling, we compute the corresponding uncertainty by evaluating the NNLO cross section with NNPDF31 NNLO PDFs obtained with $\as(m_Z)=0.119$ and $\as(m_Z)=0.117$.
Naively, one could expect relatively large effects since the process starts at ${\cal O}(\as^2)$
and small variations on $\as(m_Z)$ induce relatively large variations on $\as(m_b)$. However, it is well known that
the value of $\as(m_Z)$ and the gluon density are correlated. Correspondingly,
the ensuing effect on the NNLO cross section reported in Table~\ref{table:totalXSwithunc} is rather small and roughly independent on the collider energy.

In summary, from the results in Table~\ref{table:totalXSwithunc} we conclude that, despite the inclusion of the NNLO corrections and the consequent reduction of the scale uncertainties, the missing higher orders in the QCD perturbative expansion still represent the dominant source of theoretical uncertainty in $b{\bar b}$ production.

\begin{table}
\begin{center}
\renewcommand{\arraystretch}{1.6}
\setlength{\tabcolsep}{1em}
\begin{tabular}{|c|c|c|c|c|c|}
\hline
& $\sigma_{\rm NNLO}(\mu {\rm b})$ & $\Delta\sigma_{\rm scale}$ & $\Delta\sigma_{\rm mass}$ & $\Delta\sigma_{\rm PDFs}$ & $\Delta\sigma_{\as}$\\
\hline
$p{\bar p}$ @ 1.96~TeV & 75.4(3) & $^{+22\%}_{-21\%}$ & $^{+9.8\%}_{-8.7\%}$ & $\pm 1.3\%$ & $^{+0.9\%}_{-3.0\%}$\\
\hline
$pp$ @ 7~TeV & 288(2) & $^{+30\%}_{-24\%}$ & $^{+7.9\%}_{-7.2\%}$ & $\pm 2.8\%$ & $^{+0.3\%}_{-2.9\%}$ \\
\hline
$pp$ @ 13~TeV & 508(3) & $^{+32\%}_{-25\%}$ & $^{+7.4\%}_{-6.8\%}$ & $\pm 4.6\%$ & $^{+0.0\%}_{-3.0\%}$ \\
\hline
\end{tabular}
\end{center}
\caption{
Total cross sections and uncertainties for $b\bar{b}$ production at NNLO for $\mu_0=m_b$. The numerical errors on the last digits are stated in brackets, as in Table~\ref{table:total_vs_hathor}.
}
\label{table:totalXSwithunc}
\end{table}

\subsection{Differential distributions: Tevatron}
\label{sec:resu:tev}

We start the presentation of our differential results by showing selected distributions in $p{\bar p}$ collisions at the Tevatron ($\sqrt{s}=1.96$ TeV).
Throughout this paper we always consider observables obtained by averaging the corresponding bottom and antibottom quark distributions. In particular we consider the average of the transverse-momentum, rapidity and pseudorapidity distributions of the (anti)bottom quark, which are denoted as $p_{T,b_{\rm av}}$, $y_{b_{\mathrm{av}}}$ and $\eta_{b_{\mathrm{av}}}$, respectively.
These observables are the parton-level equivalent of the corresponding $b$-hadron distributions.

\begin{figure}[t]
\centering
\includegraphics[width=\rescalethreeplots]{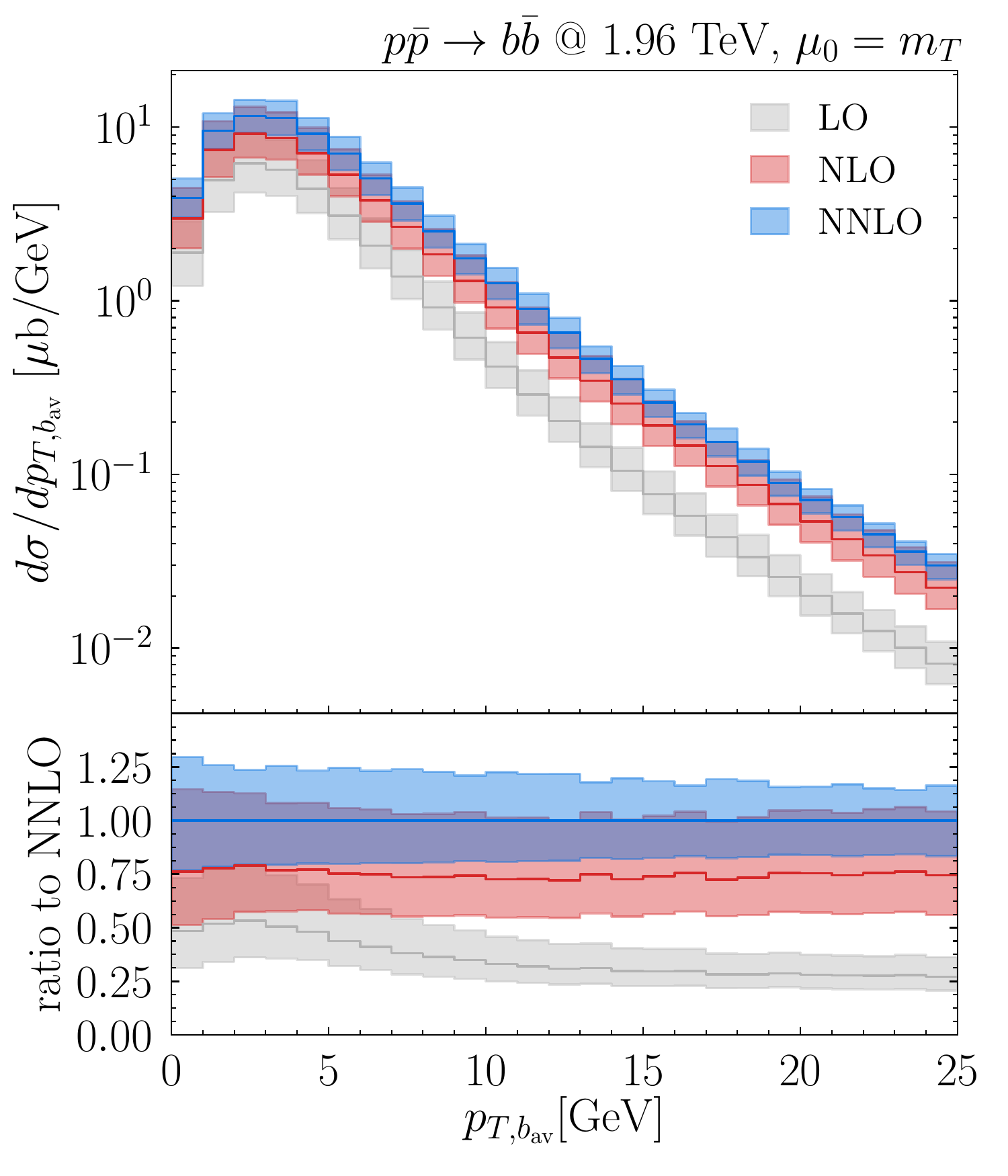}\hspacebetweenthreeplots
\includegraphics[width=\rescalethreeplots]{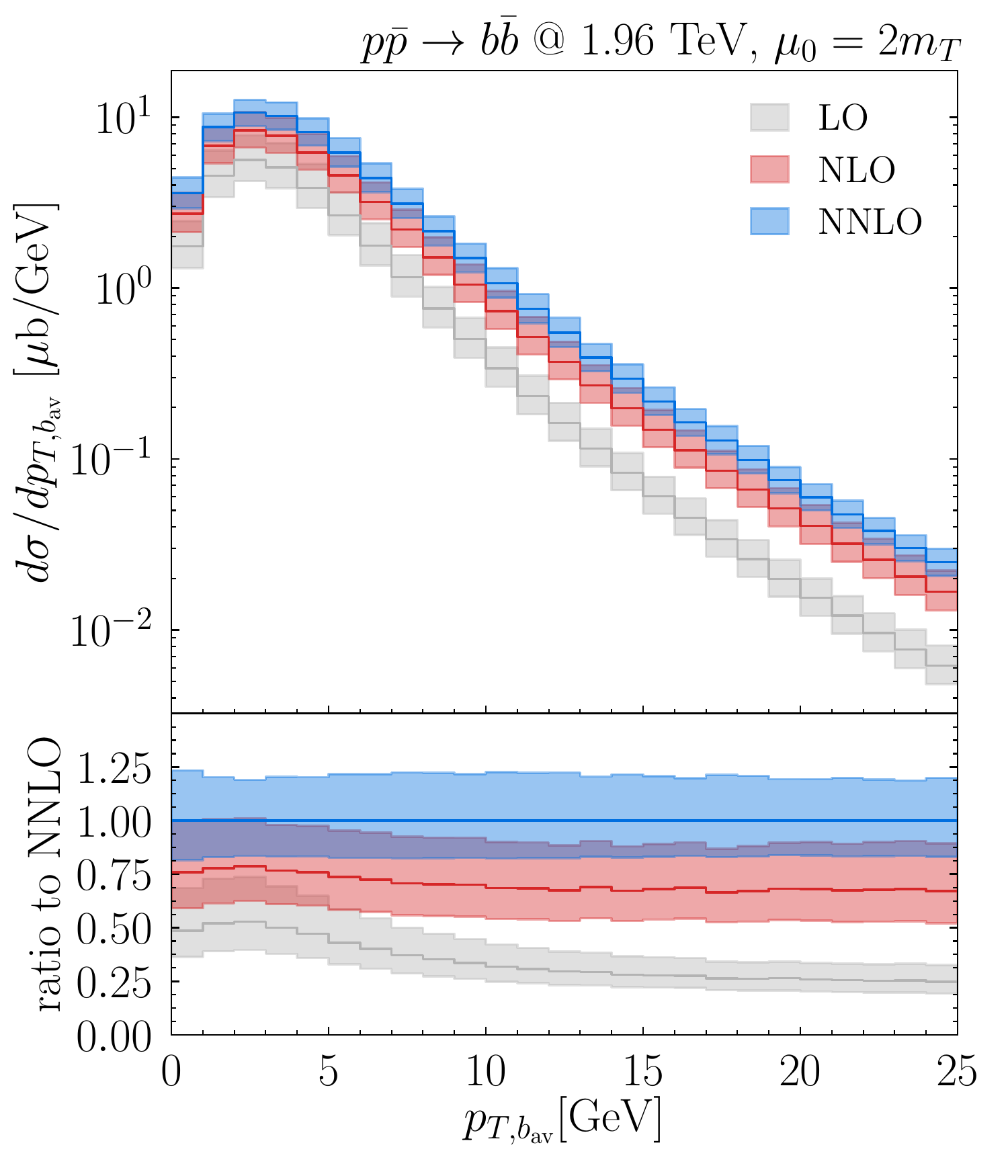}\hspacebetweenthreeplots
\includegraphics[width=\rescalethreeplots]{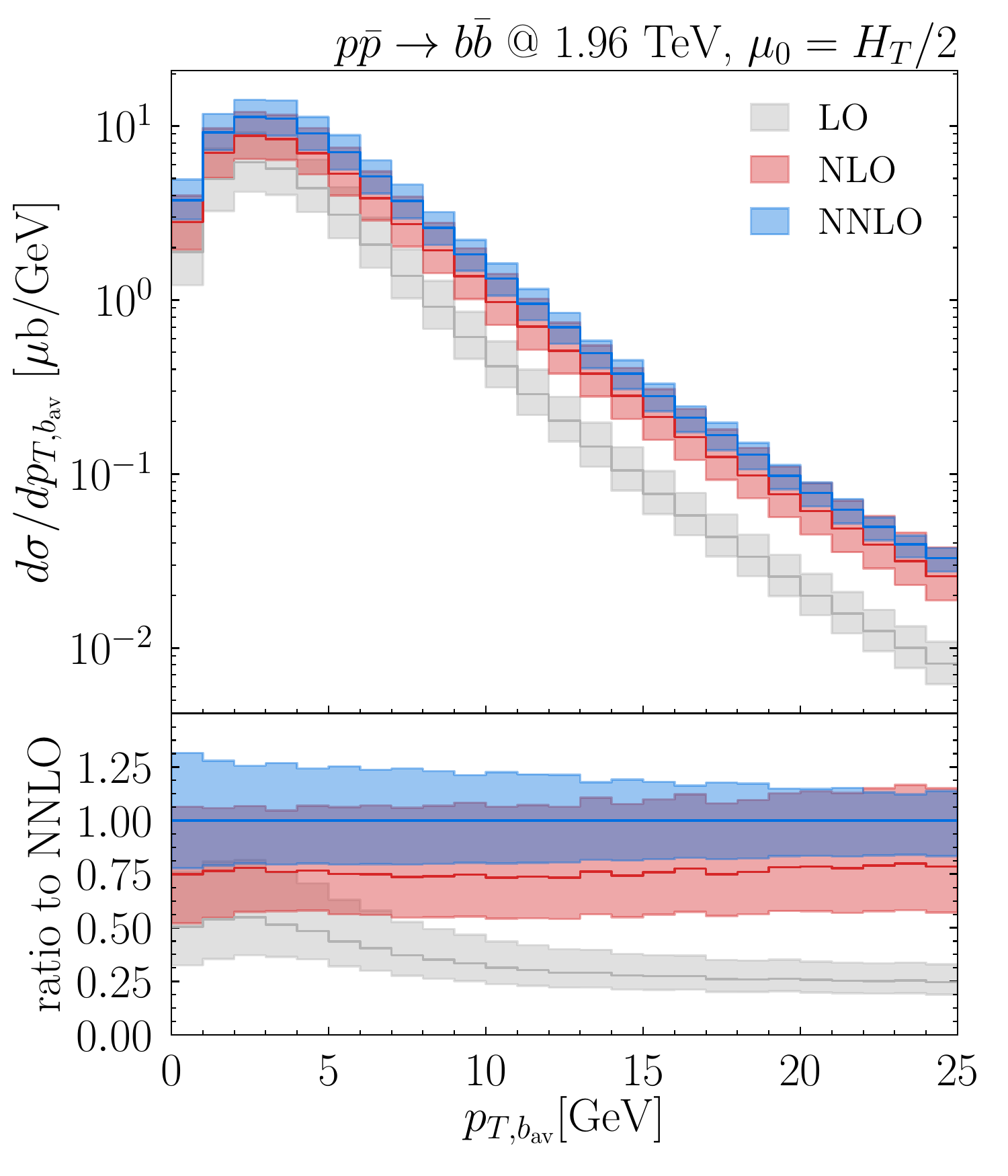}
\spaceabovefigurecaption
\caption{Transverse momentum distributions at the Tevatron, for the scale choice $\mu_0=m_T$ (left), $\mu_0=2m_T$ (central) and $\mu_0=H_T/2$ (right). The lower panels show the ratio to the NNLO predictions.}
\label{fig:pT_av.Tev}
\end{figure}

In Fig.~\ref{fig:pT_av.Tev} we present our LO, NLO and NNLO predictions for the transverse-momentum distribution.
The calculation is carried out without applying any kinematical cut on the final-state partons.
The transverse-momentum distributions of the bottom and antibottom quark are controlled by hard-scattering scales of the order of the corresponding transverse mass,
\begin{equation}
m_{T,b/\bar b}=\sqrt{m_b^2+p_{T,b/\bar b}^2}\;.
\end{equation}
Therefore, in Fig.~\ref{fig:pT_av.Tev} (left and central panels) we show predictions with the scale choices $\mu_0=m_T$ and $\mu_0=2m_T$, i.e. by using $\mu_0=m_{T,b(\bar b)}$ or $\mu_0=2m_{T,b(\bar b)}$ for the (anti)bottom transverse-momentum distribution used to
compute the average.

In addition to these two `natural' scales, we also show predictions with the dynamic scale $\mu_0=H_T/2$ (right panel in Fig.~\ref{fig:pT_av.Tev}), defined as the average of the two transverse masses:
\begin{equation}
\frac 12 H_T=\frac{m_{T,b}+m_{T,\bar b}}2\;.
\end{equation}
Note that we have $H_T/2 = m_T$ at LO.
Differential results at each order in the perturbative expansion are shown in the upper panels of Fig.~\ref{fig:pT_av.Tev}, while in the lower panels the ratio of each perturbative order to our NNLO prediction is presented.
From Fig.~\ref{fig:pT_av.Tev} we see that the distribution is peaked at $p_{T,b_{\rm av}}\sim 3$ GeV. The average transverse momentum of the (anti)bottom quark is $4.6$ GeV (at both NLO and NNLO),
i.e., as expected, quite close to the value of the bottom-quark mass. As a consequence, the total cross section receives a small contribution from the large-$p_{T,b_{\rm av}}$ region.
For instance, the region with $p_{T,b_{\rm av}}<20$ GeV gives about $99\%$ of the total cross section.

For all the considered central-scale choices LO and NLO predictions are consistent within uncertainties only in the low-$p_{T,b_{\mathrm{av}}}$ region, where the bulk of the cross section is located, while they present very different shapes in the tail of the distribution, where the NLO corrections become more sizeable. In this region the LO and NLO scale uncertainty bands do not overlap.
The inclusion of the NNLO corrections leads to a nice stabilisation of the perturbative result, analogously to what we have observed for the total cross section. In particular, the uncertainty at NNLO is smaller than at NLO, and the NLO and NNLO bands overlap in the entire region of transverse momenta.

By comparing the left and central panels in Fig.~\ref{fig:pT_av.Tev} we see that the choice of a smaller scale $\mu_0=m_T$ leads to a better overlap between the NLO and NNLO uncertainty bands, similarly to what was observed for the total cross section.
In particular, the scale choice $\mu_0=2m_T$ leads to a significantly worse perturbative convergence in the tail of the distribution.
The dynamic scale $\mu_0=H_T/2$ presents similar features to those observed for $\mu_0=m_T$, consistently with the fact that both scales are equivalent at LO, with a good overlap of the NLO and NNLO bands for all values of $p_{T, b_\mathrm{av}}$.

Perturbative predictions for bottom-quark production can be compared to experimental data for the inclusive production of $b$-hadrons.
A precise comparison in the region of large transverse momenta of the bottom quark (i.e. $p_T\gg m_b$)
would require the resummation of the logarithmically enhanced terms at large transverse momenta~\cite{Cacciari:1993mq}.
Such higher-order contributions are included up to next-to-leading logarithmic accuracy in the resummed predictions
of Refs.~\cite{Cacciari:1998it,Cacciari:2001td}. The non-perturbative effects of the fragmentation of the bottom quark into the triggered $b$-hadron
should eventually be accounted for by folding the perturbative result with an appropriate non-perturbative fragmentation function.
In this paper we limit ourselves to considering perturbative predictions up to NNLO, and a thorough comparison with experimental data is beyond the scope of this work.
In \ref{app:FONLL} we present a comparison of our NLO and NNLO results with the
FONLL prediction from Ref.~\cite{webpage}.
Such comparison shows that in the region of transverse momenta considered in this paper the resummation effects have a limited impact on the NLO transverse-momentum and (pseudo)rapidity distributions, and
that our NNLO results have smaller perturbative uncertainties than the FONLL results and can thus be considered more accurate.
The impact of non-perturbative fragmentation is typically rather small on $p_T$-inclusive observables, definitely smaller than the NNLO perturbative uncertainties.

Having in mind the above issues and limitations, our fixed-order perturbative predictions can be compared with experimental measurements for inclusive $b$-hadron production.

\begin{figure}[t]
\centering
\includegraphics[width=\rescalethreeplots]{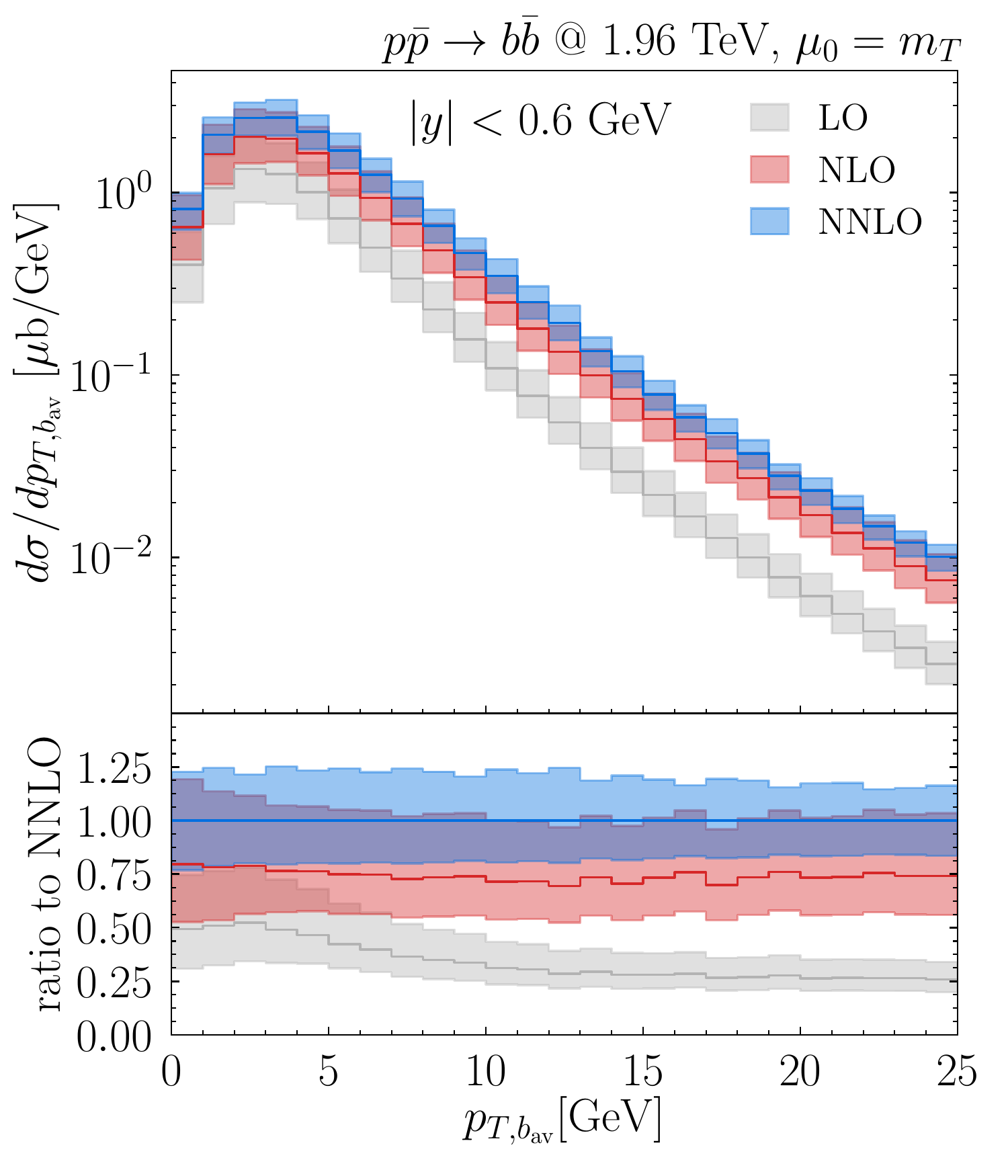}\hspacebetweenthreeplots
\includegraphics[width=\rescalethreeplots]{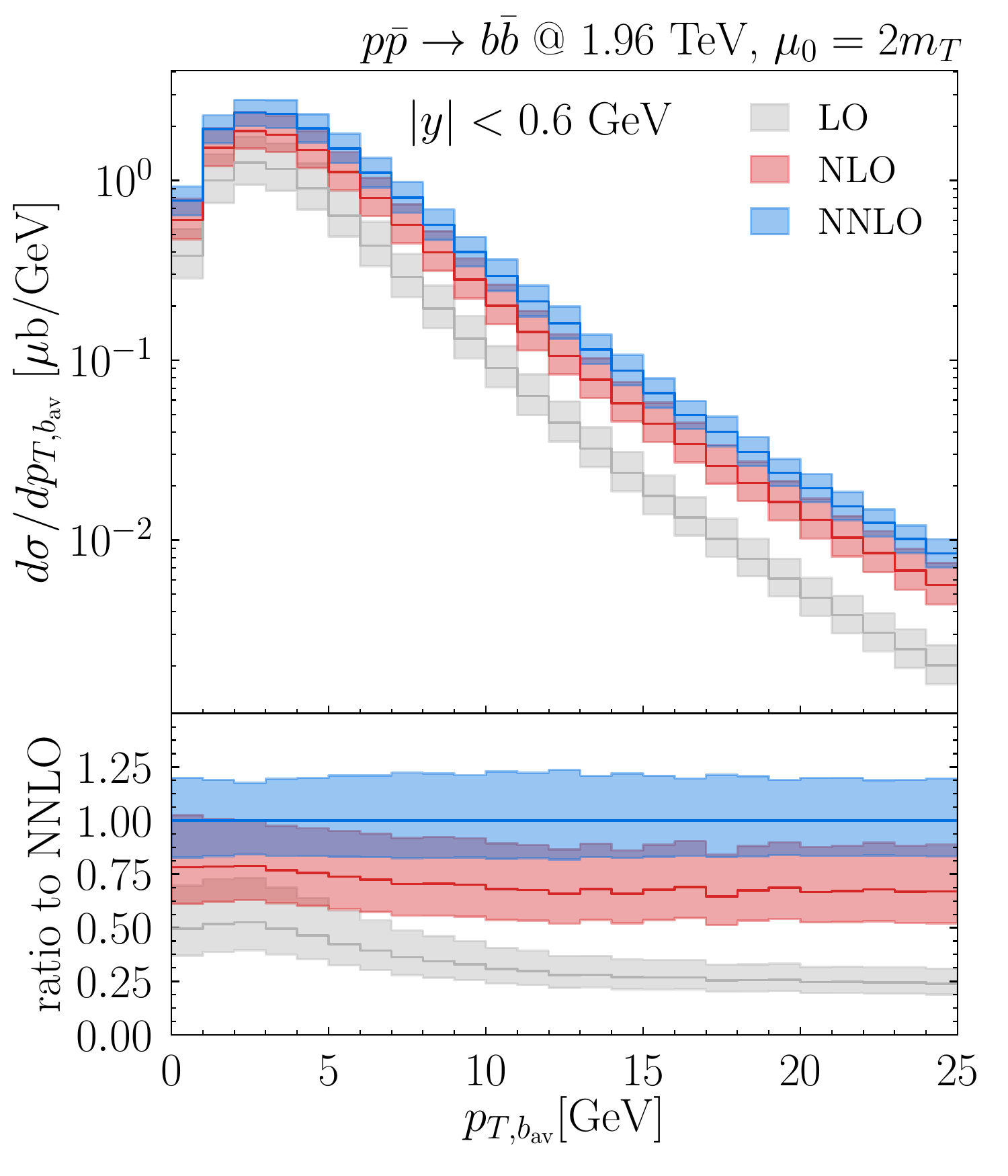}\hspacebetweenthreeplots
\includegraphics[width=\rescalethreeplots]{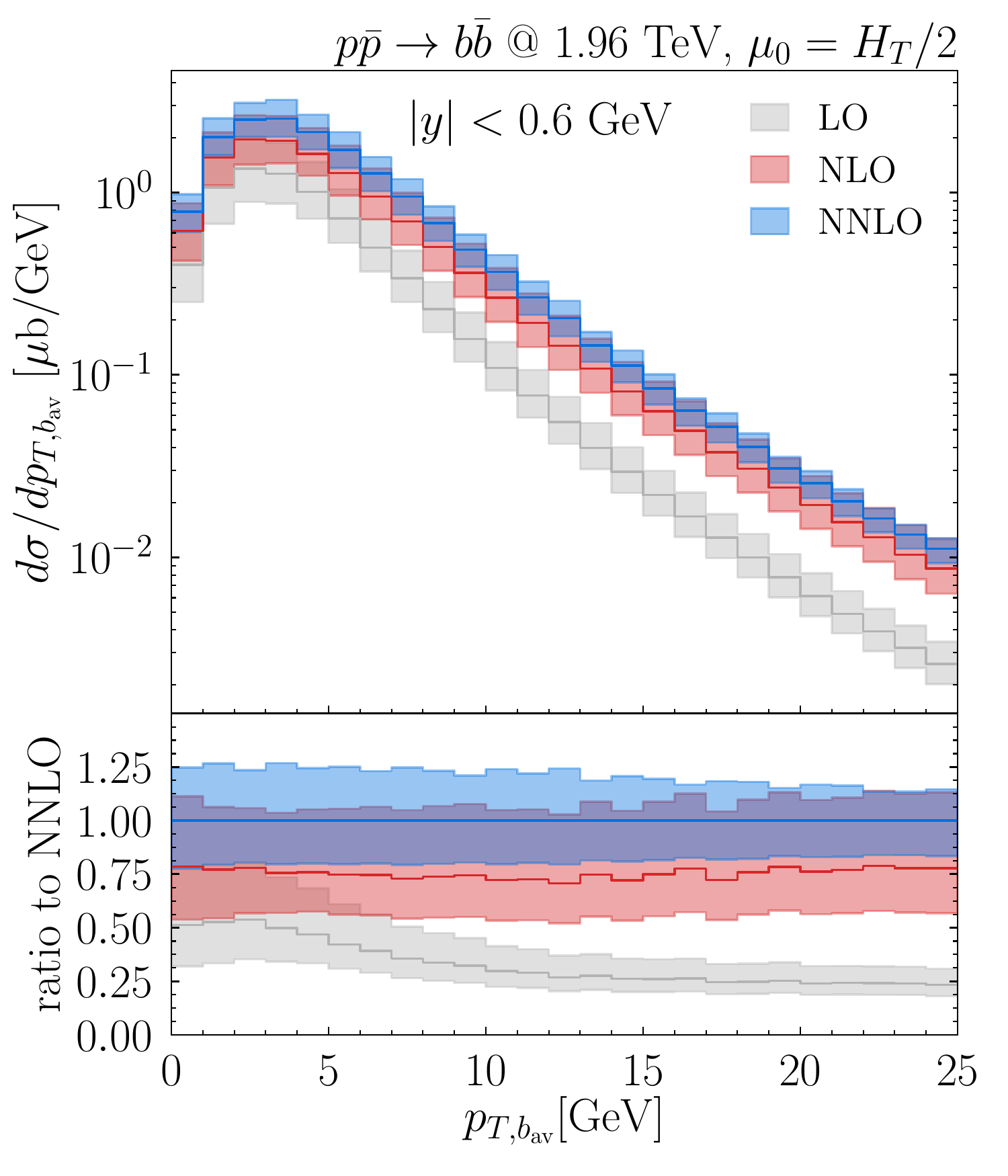}
\spaceabovefigurecaption
\caption{Transverse momentum distributions at the Tevatron with the rapidity cut $|y|<0.6$, for the scale choice $\mu_0=m_T$ (left), $\mu_0=2m_T$ (central) and $\mu_0=H_T/2$ (right). The lower panels show the ratio to the NNLO predictions.}
\label{fig:rap.pT_av.Tev}
\end{figure}

In Ref.~\cite{Acosta:2004yw}, the CDF collaboration performed a measurement of the $J/\psi$ cross section in the rapidity range $|y|<0.6$.
The total $b$-hadron production cross section in the same rapidity region was extracted by using a Monte Carlo simulation of the decay kinematics of $b$-hadrons to the final states containing a $J/\psi$ hadron.
The result is
\begin{equation}\label{eq:tot_tev}
\sigma^{H_b} (|y_{H_b}|<0.6)=17.6\pm 0.4(\text{stat})^{+2.5}_{-2.3}(\text{syst})\, \mu \text{b}\, .
\end{equation}
In order to study the possible effect of a rapidity cut in our parton-level calculation,
we have repeated the computation of the transverse-momentum distribution reported in Fig.~\ref{fig:pT_av.Tev} by applying the constraint $|y_{b/\bar{b}}|<0.6$, see Fig.~\ref{fig:rap.pT_av.Tev}.%
\footnote{The rapidity cut is applied only to the corresponding parton (bottom or antibottom) before computing the average between the two transverse momentum distributions.}
%%%%%%%%%%%%%%%%%
%%%%%%%%%%%%%%%%%
By comparing the results in Figs.~\ref{fig:pT_av.Tev} and \ref{fig:rap.pT_av.Tev} we observe that the inclusion of the rapidity cut does not significantly modify the behaviour of the perturbative series, and that the shapes of the distributions remains rather similar.

We have computed the parton-level analogue of the $b$-hadron cross section in Eq.~(\ref{eq:tot_tev}).
In order to do so, we compute two independent cross sections, with cuts on the rapidity of the bottom \emph{or} the antibottom quark, respectively. These two cross sections are then averaged.
Since the characteristic scale for this observable is the bottom-quark mass $m_b$, we have chosen $\mu_0=m_b$ as the central scale.
We find
\begin{align}
\sigma^{b/\bar b}_{\text{LO}} (|y_{b/\bar{b}}|<0.6) &=  7.840(3) ^{+51\%}_{-34\%}\;\mu\text{b,}\nn\\
\sigma^{b/\bar b}_{\text{NLO}} (|y_{b/\bar{b}}|<0.6) &=  14.282(6) ^{+53\%}_{-28\%}\;\mu\text{b,}\\
\sigma^{b/\bar b}_{\text{NNLO}} (|y_{b/\bar{b}}|<0.6) &=  17.87(12) ^{+22\%}_{-21\%}\;\mu\text{b}\nn
\end{align}
and compare these results with the inclusive predictions presented in Table~\ref{table:totalXS}. The cross section turns out to be about 4 times smaller in the presence of this rapidity cut.
The NLO and NNLO $K$-factors are close to those for the total cross section, and also the scale uncertainties are rather similar.
This is a consequence of the fact that the impact of QCD radiative corrections is rather uniform in rapidity (see e.g. Fig.~\ref{fig:y_av.LHC}).
Comparing our perturbative predictions with the CDF measurement in Eq.~(\ref{eq:tot_tev}), we find that
NNLO corrections considerably improve the agreement with the data. Scale uncertainties are, although still sizeable, largely reduced at NNLO,
and only at this order they approach the size of the experimental uncertainties.

\subsection{Differential distributions: LHC}
\label{sec:resu:lhc}

% pt and rapidity distributions

\begin{figure}[t]
\centering
\includegraphics[width=\rescaletwoplots]{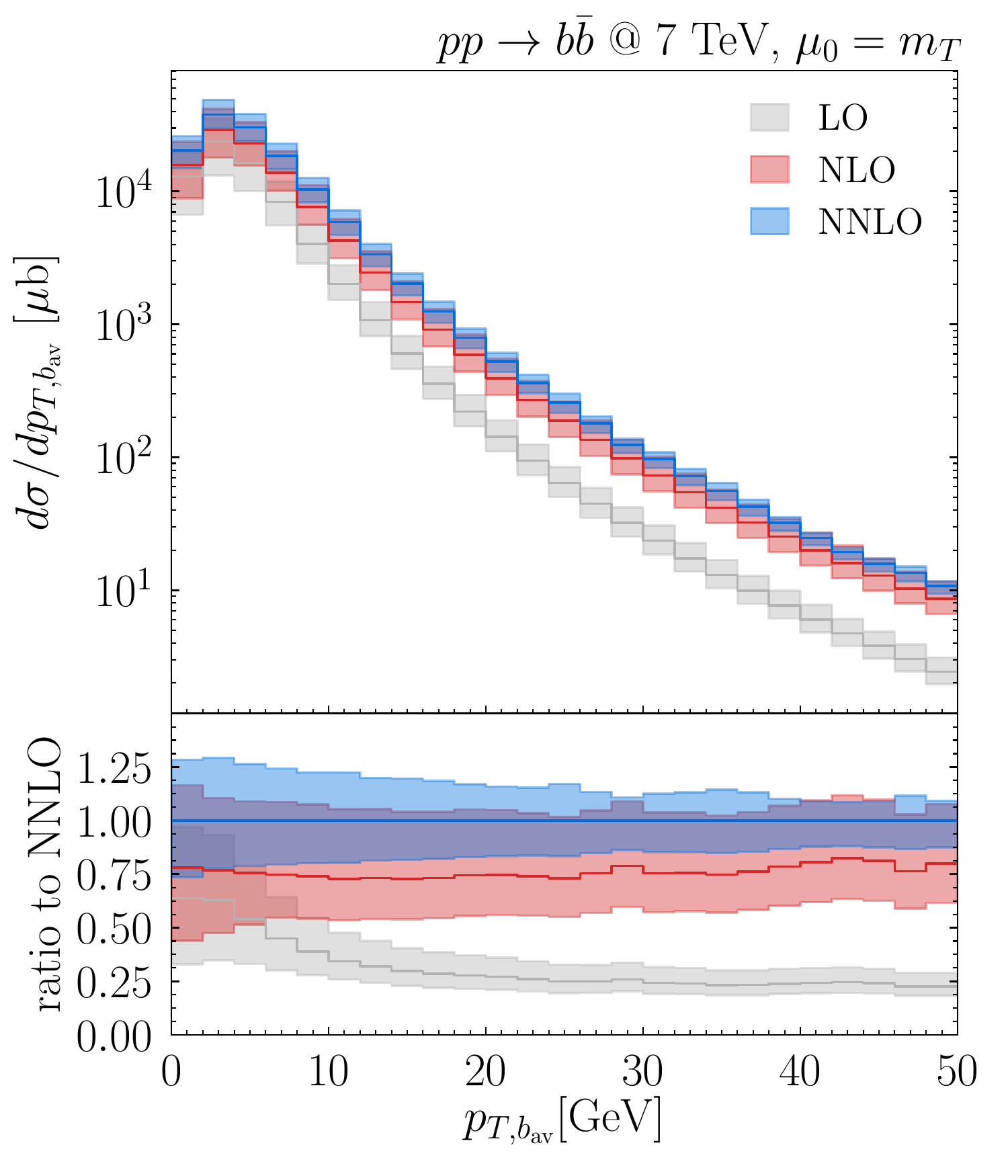}\hspacebetweentwoplots
\includegraphics[width=\rescaletwoplots]{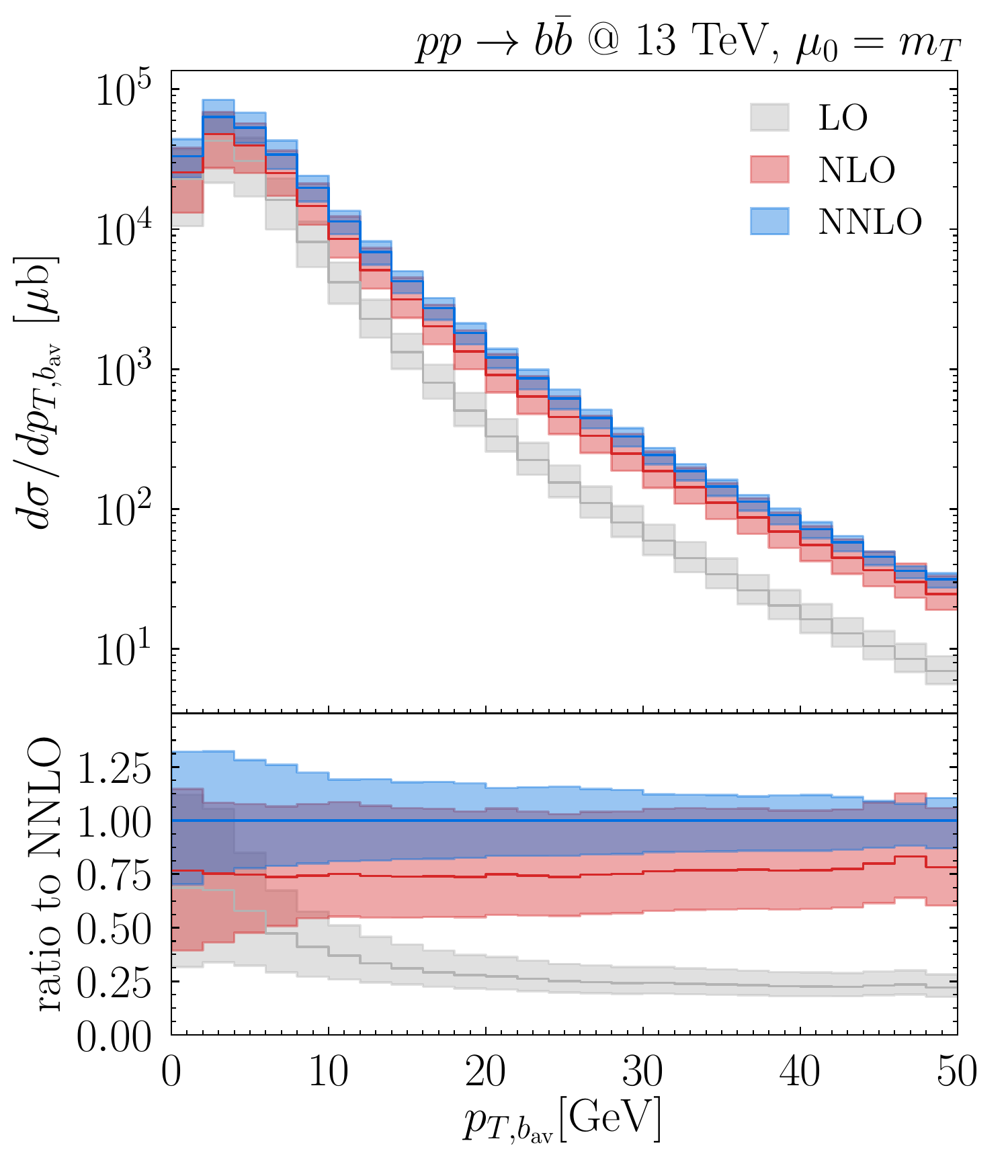}
\spaceabovefigurecaption
\caption{Transverse momentum distribution at the LHC for the scale choice $\mu_0=m_T$, for $\sqrt s = 7$~TeV (left) and $13$~TeV (right). The lower panels show the ratio to the NNLO prediction.}
\label{fig:pT_av.LHC}
\end{figure}

We start the presentation of our differential results for the LHC by considering
the transverse momentum distribution of the bottom quark. Using different central scales, such as $\mu_0=m_T$, $\mu_0=2m_T$ and $\mu_0=H_T/2$, we obtain relative differences that are similar to those
predicted at the Tevatron (Fig.~\ref{fig:pT_av.Tev}). Therefore, we set $\mu_0=m_T$, and in Fig.~\ref{fig:pT_av.LHC} we show our LHC results at $\sqrt s = 7$~TeV (left panel) and $13$~TeV (right panel).
The shape of the $p_{T,b_\mathrm{av}}$ distribution is rather similar to what was observed at the Tevatron, and the peak is located at $p_{T,b_\mathrm{av}}\sim 4$ GeV.
The average transverse momentum is only slightly larger at LHC energies, being (at both NLO and NNLO) 5.5 GeV and 5.9 GeV at $\sqrt s = 7$ and $13$~TeV, respectively.

We see that, as already observed at the Tevatron, LO and NLO predictions are consistent within uncertainties only in the low-$p_{T,b_{\mathrm{av}}}$ region, while they present very different shapes in the tail of the distribution, where the NLO corrections become very large. In this region the LO and NLO scale uncertainty bands do not overlap.
The inclusion of NNLO corrections leads to a nice stabilisation of the perturbative result, analogously to what we have observed for the total cross section. In particular, the uncertainty band at NNLO is smaller than at NLO, and it overlaps with the latter over the entire region of transverse momenta.
At small transverse momenta the NNLO scale uncertainty is larger than at the Tevatron, consistently with our observation for the corresponding total cross sections. On the contrary, at large transverse momenta the NNLO band is smaller at the LHC (note that the plots in Fig.~\ref{fig:pT_av.LHC} extend to $p_{T,b_\mathrm{av}}=50$~GeV, while the Tevatron result in Fig.~\ref{fig:pT_av.Tev} is shown up to $25$~GeV).

%%%%%%%%%%%%%%%%%
\begin{figure}[p]
\centering
\includegraphics[width=\rescaletwoplots]{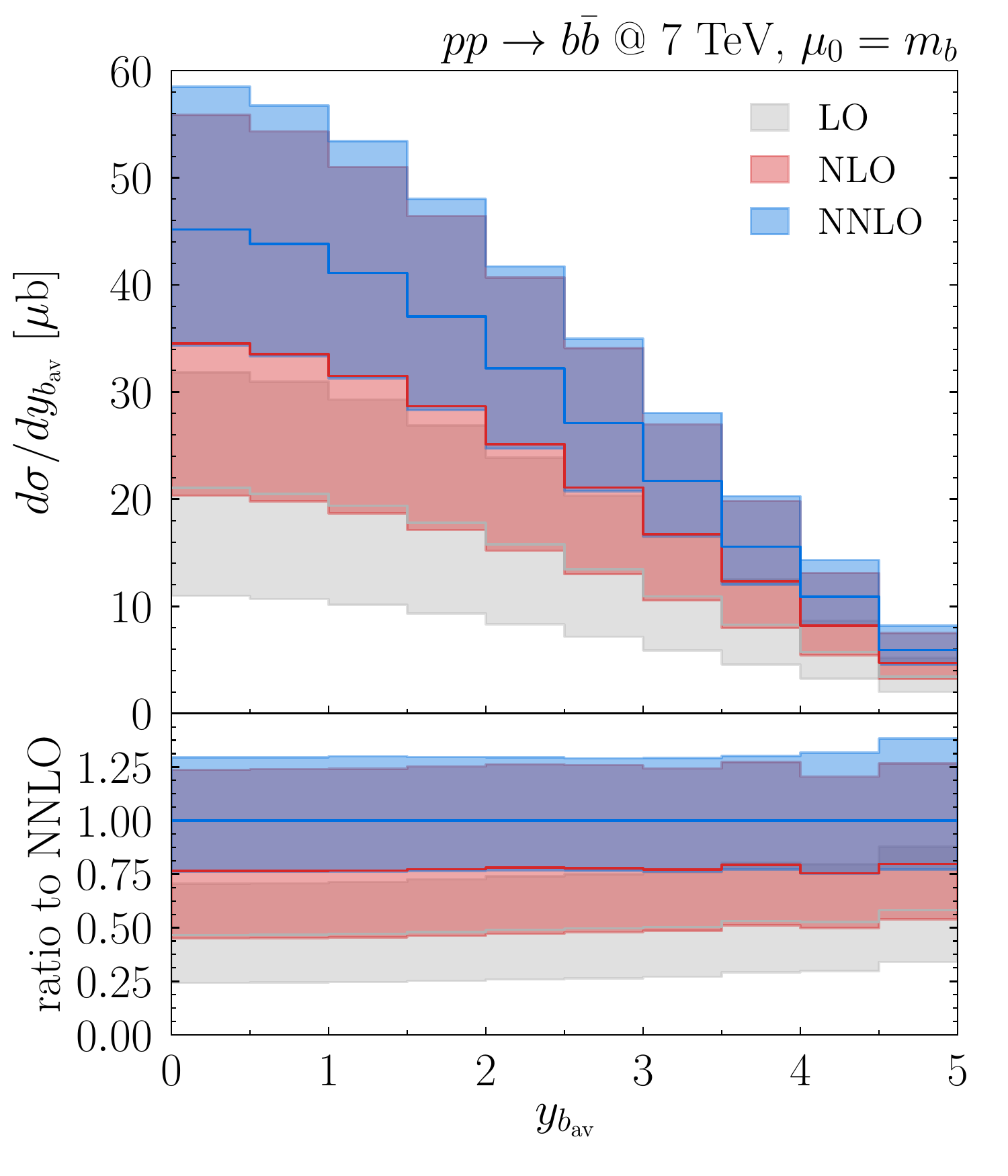}\hspacebetweentwoplots
\includegraphics[width=\rescaletwoplots]{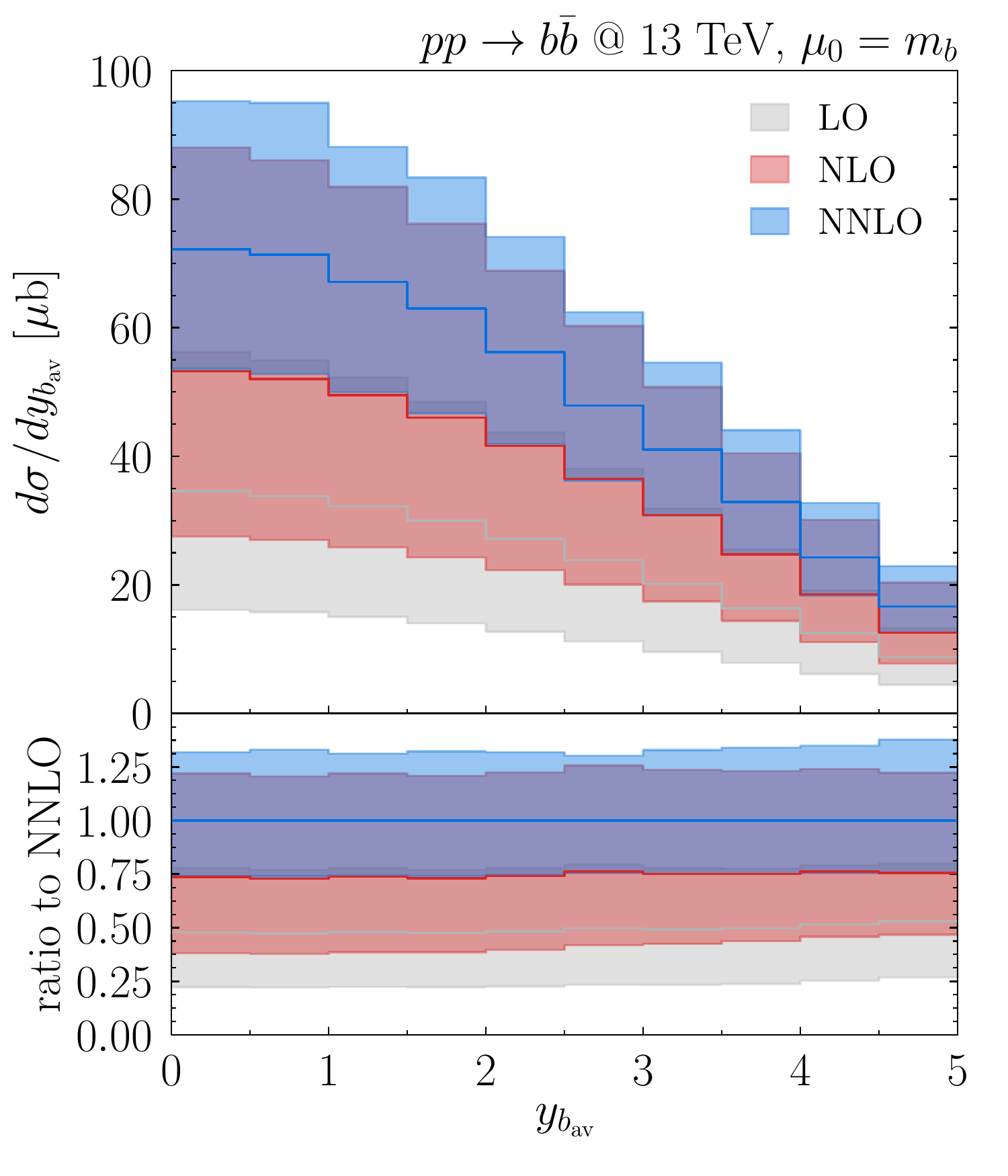}
\spaceabovefigurecaption
\caption{Rapidity distribution of the bottom quark at the LHC for $\sqrt s = 7$~TeV (left) and $13$~TeV (right). The lower panels show the ratio to the NNLO prediction.}
\label{fig:y_av.LHC}
%\end{figure}
\vspace*{5ex}
%
%\begin{figure}[t]
\centering
\includegraphics[width=\rescaletwoplots]{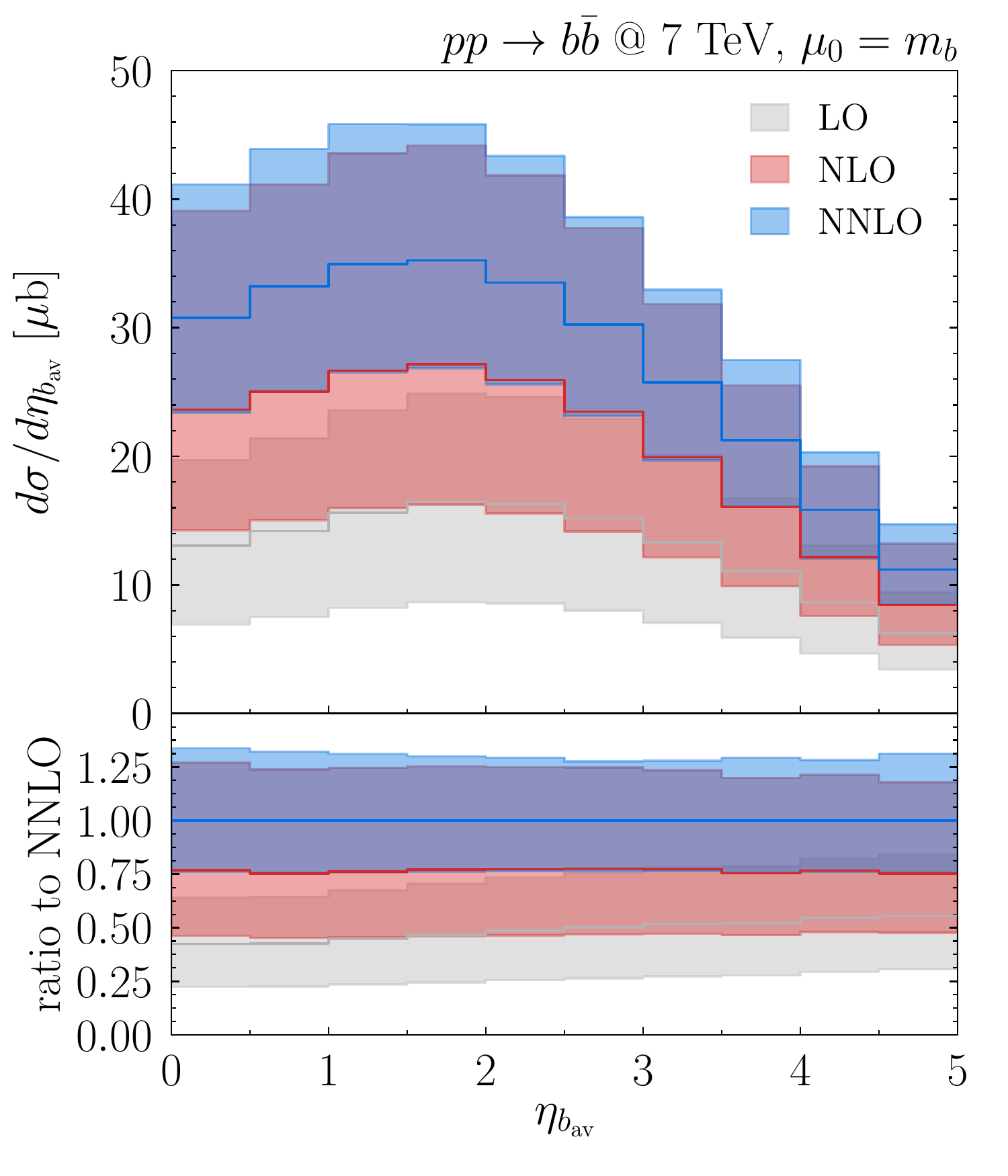}\hspacebetweentwoplots
\includegraphics[width=\rescaletwoplots]{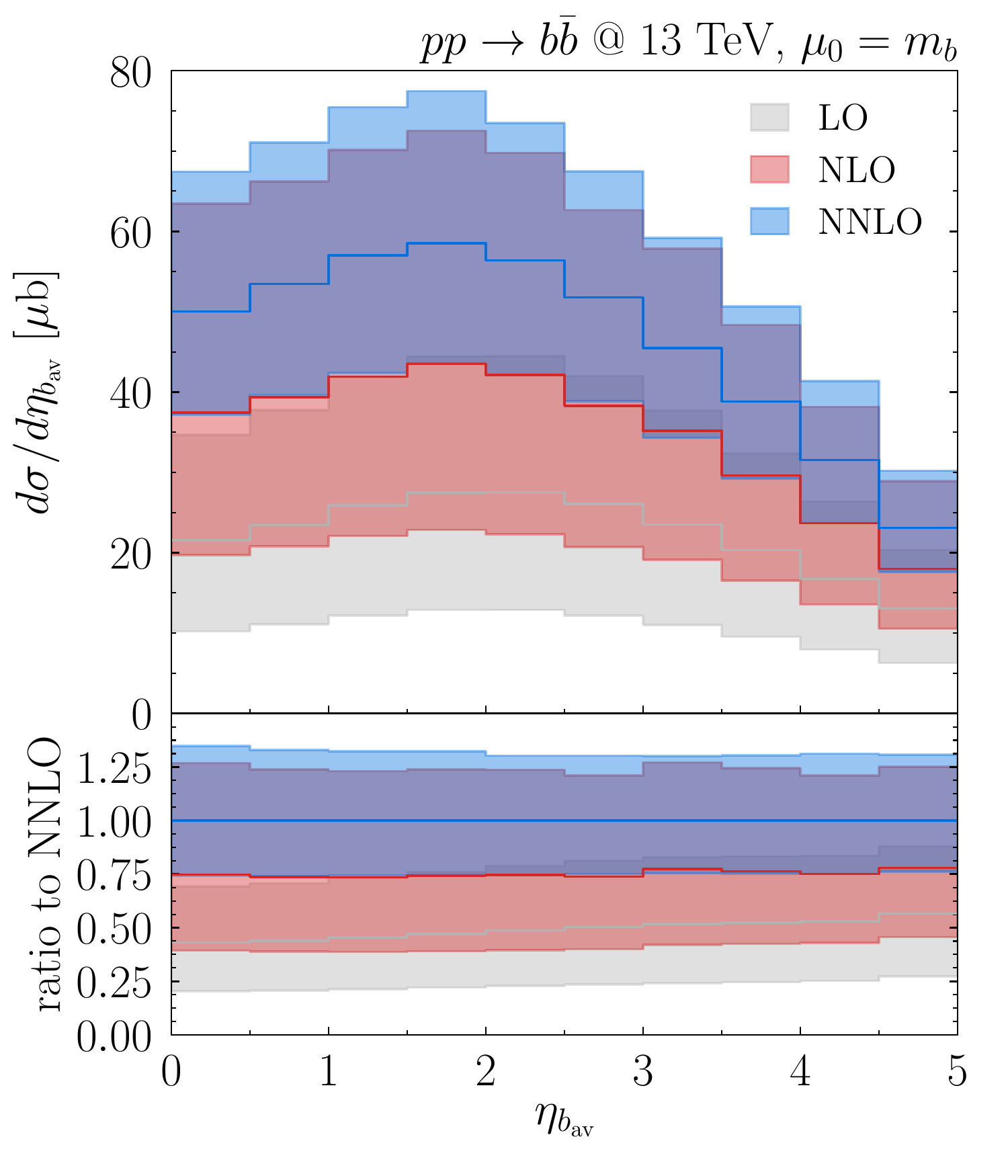}
\spaceabovefigurecaption
\caption{Pseudorapidity distribution of the bottom quark at the LHC for $\sqrt s = 7$~TeV (left) and $13$~TeV (right). The lower panels show the ratio to the NNLO prediction.}
\label{fig:neweta_av.LHC}
\end{figure}
%%%%%%%%%%%%%%%%%
The rapidity distribution of the bottom quark, computed with $\mu_0=m_b$, is shown in Fig.~\ref{fig:y_av.LHC} for $\sqrt s = 7$~TeV (left) and $13$~TeV (right).
The two distributions at different LHC energies show a similar behaviour. The impact of QCD radiative corrections is uniform in rapidity and, therefore,
it does not produce sizeable changes in the shape\footnote{The shape of
$d\sigma/dy$ at different perturbative orders also depends on the PDFs
at the corresponding order.
In general, shape differences of the PDFs at different orders can
produce ensuing shape variations of $d\sigma/dy$.}
of the distribution (even from LO to NLO).
Since the radiative corrections are relatively flat, they are of the same size as those for the total cross section.
The NNLO results almost completely overlap with the NLO results, and
they show smaller scale uncertainties, which are about $\pm 30\%$ over the whole spectrum.

% pseudorapidity distribution

In Fig.~\ref{fig:neweta_av.LHC} we report analogous results to those of
Fig.~\ref{fig:y_av.LHC}, but we consider the pseudorapidity ($\eta$) distribution of the bottom quark rather than its rapidity distribution. The most striking effect that we observe in going from Fig.~\ref{fig:y_av.LHC} to Fig.~\ref{fig:neweta_av.LHC}
is the different shape (independently of the perturbative order) of the rapidity and pseudorapidity distributions. The rapidity cross section is maximal at $y=0$ and monotonically decreases as $y$ increases. The pseudorapidity cross section has a maximum at a finite (non-vanishing) value of $\eta$ and a local minumum at $\eta=0$,
where it shows a `central-pseudorapidity dip' (the value of $d\sigma/d\eta$ at
$\eta=0$ is smaller than the value of $d\sigma/dy$ at $y=0$).
These shape differences are a well-known fact: they have a general kinematical origin
and are due to the non-vanishing mass of the produced particle. We recall the origin of these shape differences in \ref{app:eta-y}.

Examining the NLO and NNLO radiative corrections, we note that they have a highly similar effect (at both the qualitative and the quantitative levels) on the rapidity and pseudorapidity distributions of the $b$ quark (see the ratio plots in
Figs.~\ref{fig:y_av.LHC} and~\ref{fig:neweta_av.LHC}). This feature is a consequence of the fact that $d\sigma/dy$ and $d\sigma/d\eta$ mainly probe the same underlying dynamics, and that their relative differences are basically of kinematical origin (see the discussion in \ref{app:eta-y}).

%%%%%%%%%%%%%%
% normalised %
%%%%%%%%%%%%%%

Despite the inclusion of NNLO corrections, the results in Figs.~\ref{fig:pT_av.LHC} and \ref{fig:y_av.LHC} show that perturbative uncertainties in transverse-momentum and rapidity distributions are still relatively large.
A possible attempt to reduce scale uncertainties is to consider {\it normalised} distributions, as discussed in the following.

%%%%%%%%%%%%%%%%%
\begin{figure}[t]
\centering
\includegraphics[width=\rescaletwoplots]{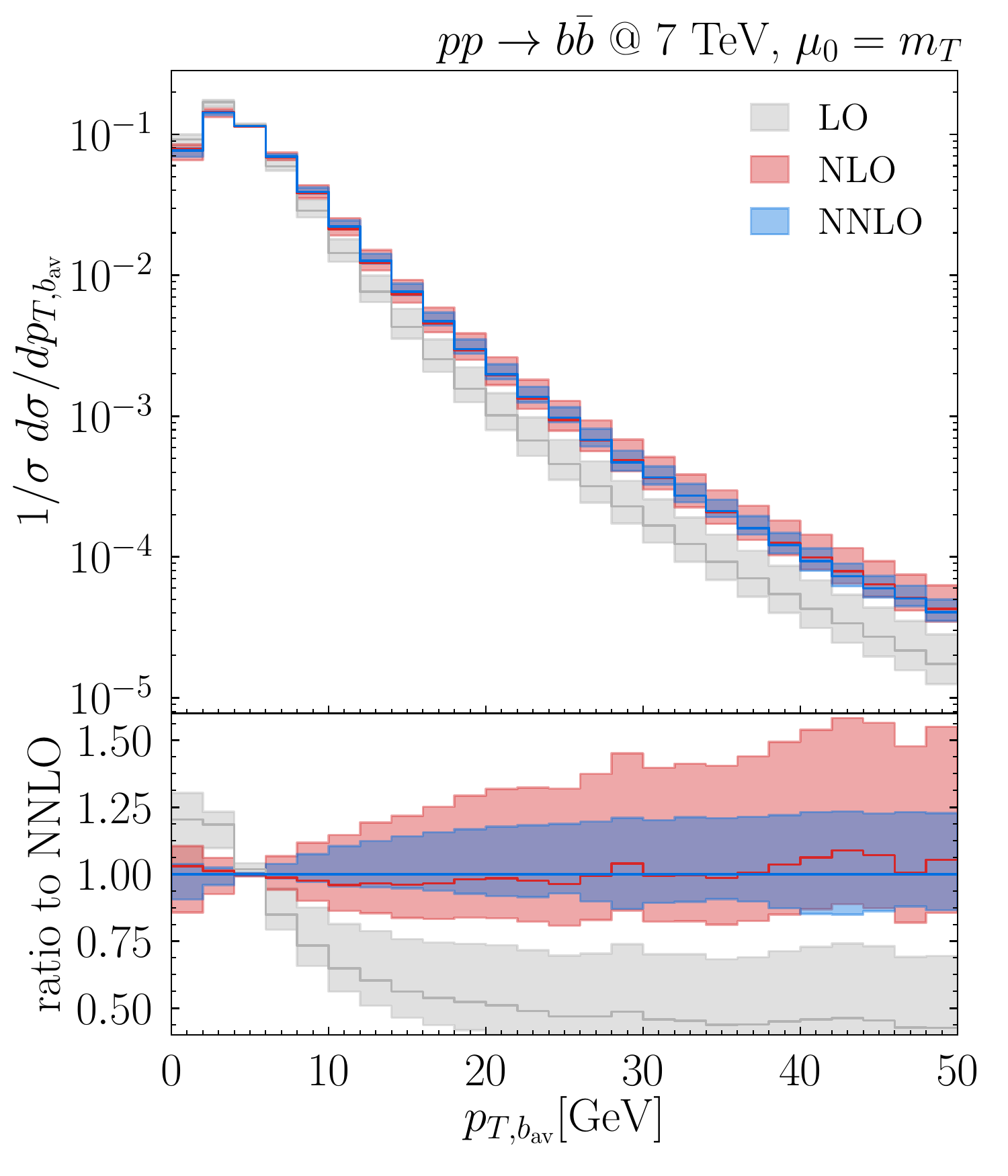}\hspacebetweentwoplots
\includegraphics[width=\rescaletwoplots]{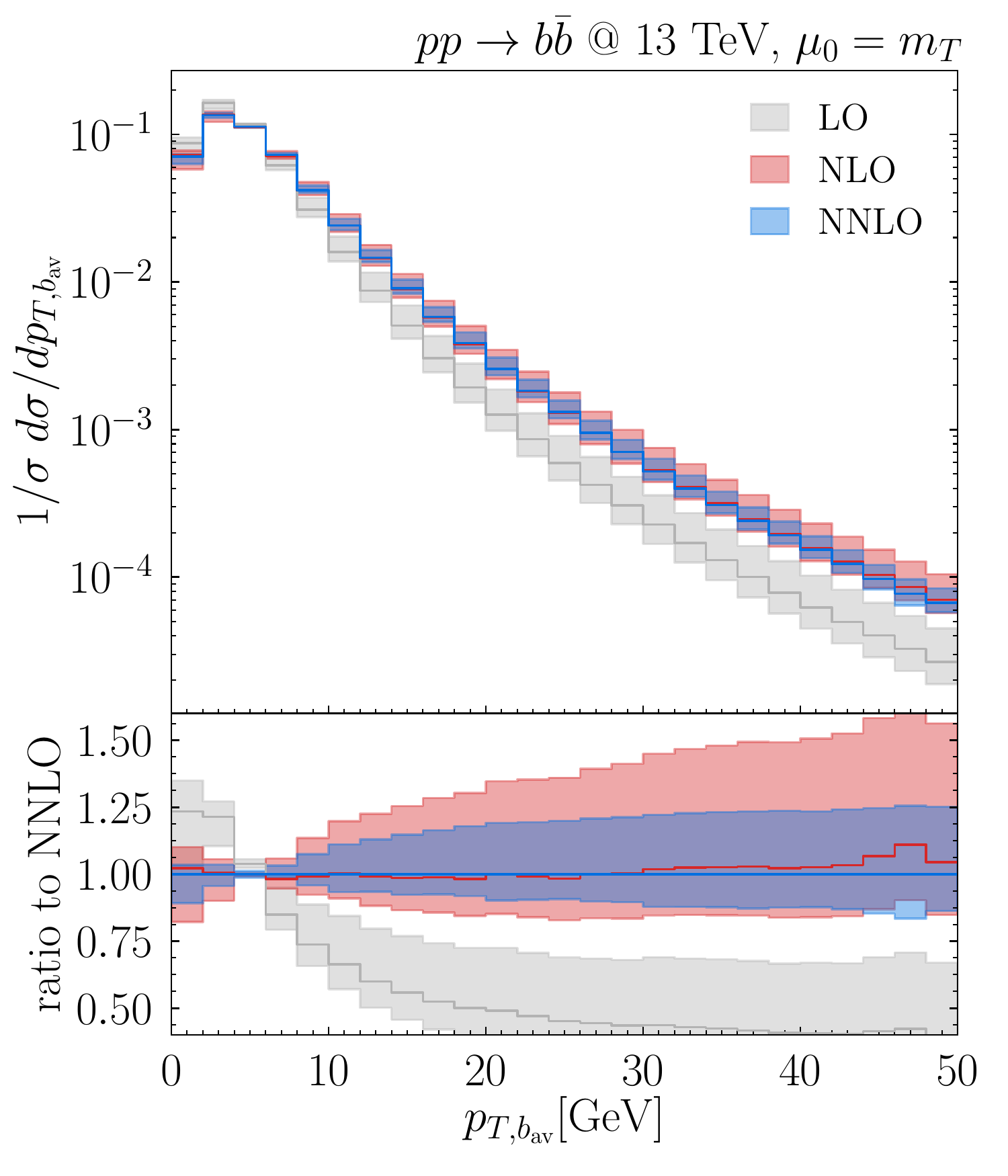}
\spaceabovefigurecaption
\caption{Normalised transverse momentum distribution of the bottom quark at the LHC for $\sqrt s = 7$~TeV (left) and $13$~TeV (right). The lower panels show the ratio to the NNLO prediction.}
\label{fig:pT_av.LHC.norm}
\end{figure}
%%%%%%%%%%%%%%%%%
In Fig.~\ref{fig:pT_av.LHC.norm} we consider the normalised $p_{T,b_\mathrm{av}}$ distribution at LO, NLO and NNLO.
The scale uncertainty bands are obtained as follows: for each scale choice needed to study the 7-point scale variation the corresponding distribution is normalised to unity, and the envelope of each
normalised distribution is constructed. The scale uncertainties in the peak region are strongly reduced, whereas at high transverse momenta the NLO and NNLO uncertainty bands are slightly larger than
those observed in Fig.~\ref{fig:pT_av.LHC}. This is not unexpected,
since the low-$p_{T,b_\mathrm{av}}$ region gives the dominant contribution to the total cross section.
Therefore, the scale uncertainties of the total and differential cross sections are strongly correlated at low $p_{T,b_\mathrm{av}}$, and increasingly decorrelated as $p_{T,b_\mathrm{av}}$ increases.

%%%%%%%%%%%%%%%%%
\begin{figure}[t]
\centering
\includegraphics[width=\rescaletwoplots]{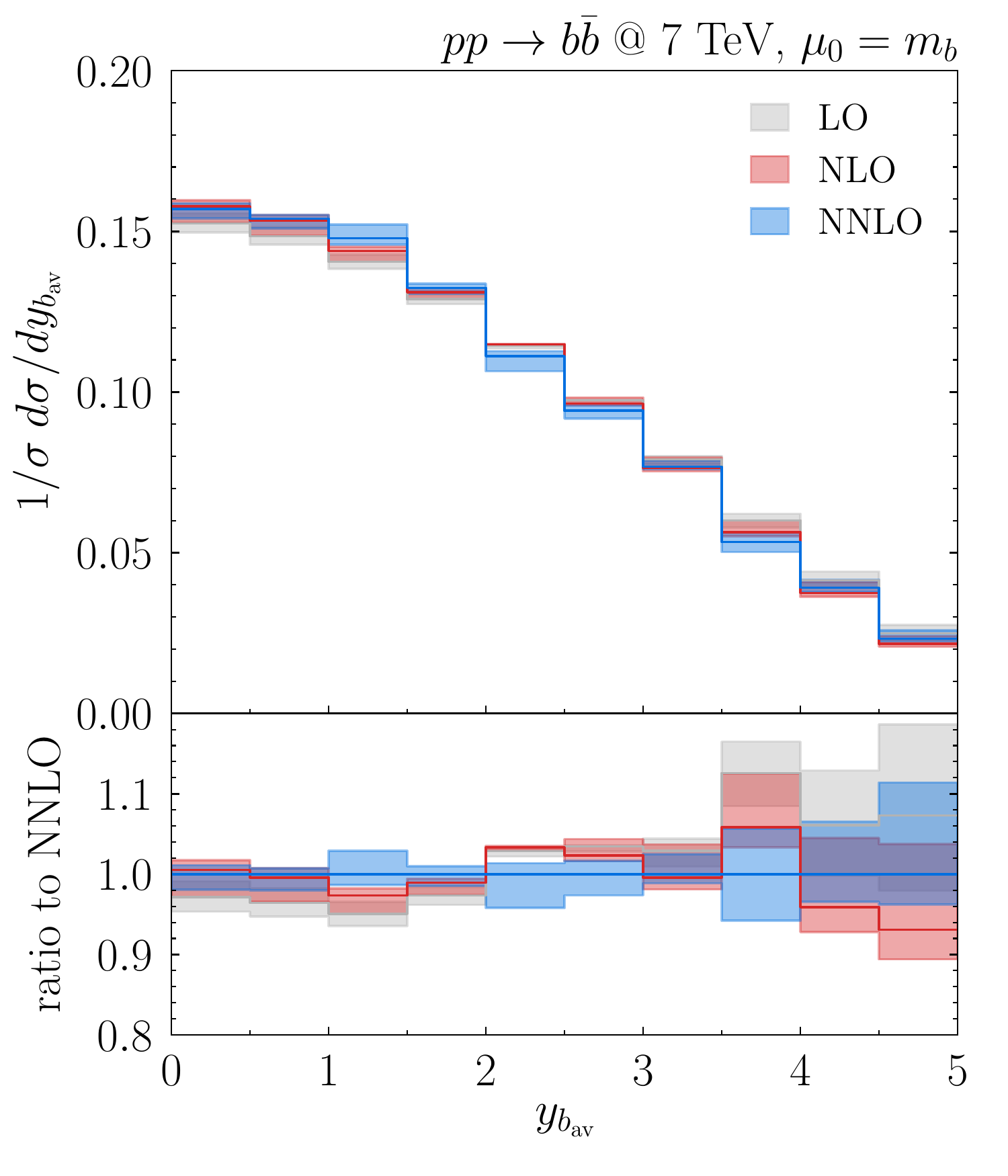}\hspacebetweentwoplots
\includegraphics[width=\rescaletwoplots]{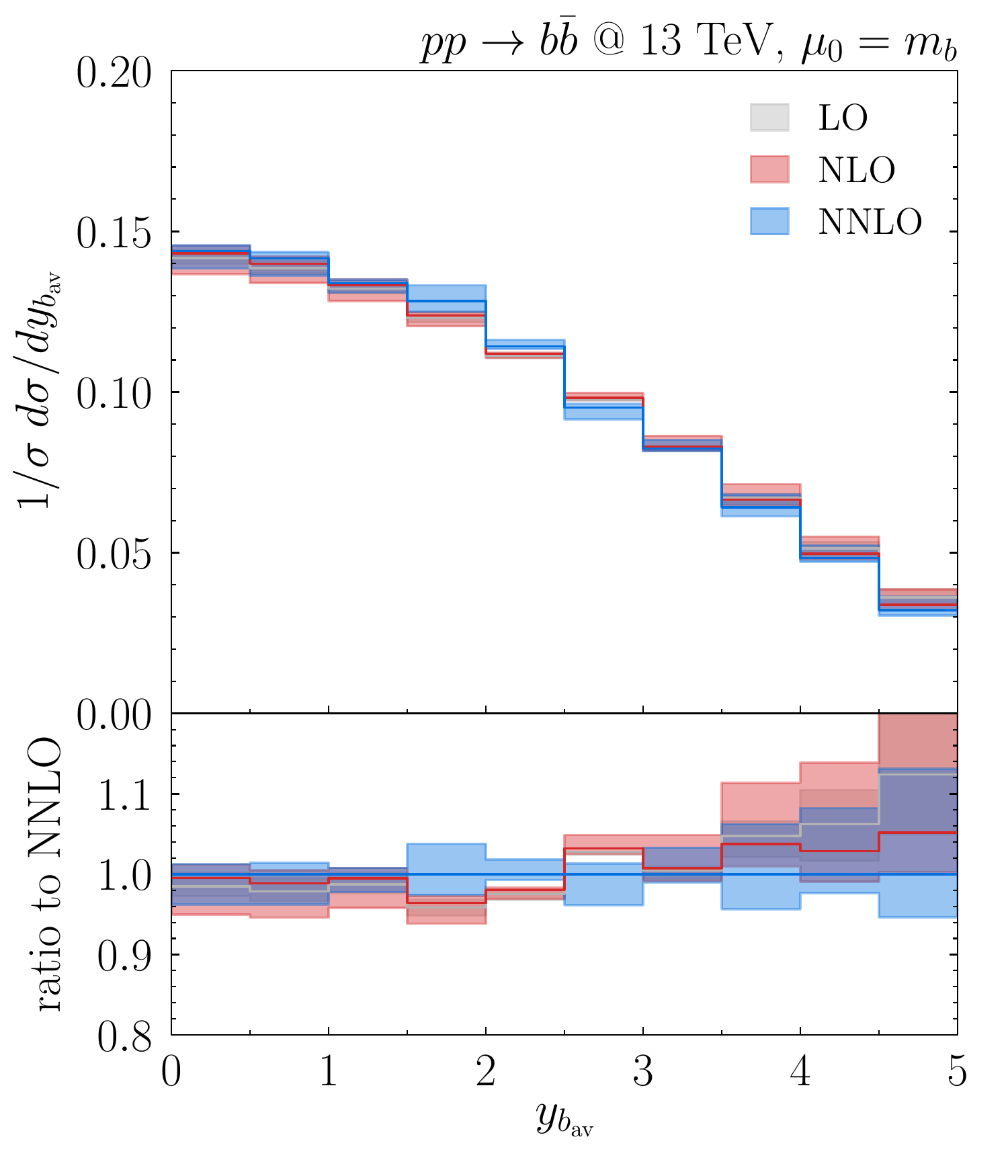}
\spaceabovefigurecaption
\caption{Normalised rapidity distribution at the LHC for $\sqrt s = 7$~TeV (left) and $13$~TeV (right). The lower panels show the ratio to the NNLO prediction.}
\label{fig:y_av.LHC.norm}
\end{figure}
%%%%%%%%%%%%%%%%%
We repeat the same procedure for the rapidity distribution. In Fig.~\ref{fig:y_av.LHC.norm} we show the normalised rapidity distribution of the bottom quark,
which is constructed in the same manner as for Fig.~\ref{fig:pT_av.LHC.norm}.
In this case, since the impact of QCD radiative corrections is rather uniform, the normalised distribution is quite stable and shows reduced scale uncertainties, except for the large-rapidity region.

%%%%%%%%%%
% ratios %
%%%%%%%%%%

An alternative strategy to reduce theoretical uncertainties is to consider ratios between distributions computed at different collider energies.
In the context of heavy-quark production, the use of such ratios was proposed in Ref.~\cite{Cacciari:2015fta}, to the purpose of constraining the gluon PDF.
Assuming that the scale variations at different energies are fully correlated (i.e.\ the same values of the renormalisation and factorisation scales can be used in the numerator and the denominator when constructing the ratio), the ratios of differential distributions at different energies exhibit a smaller sensitivity to scale variations.
The validity of this assumption can be tested by comparing the ratios at different perturbative orders, and checking the reliability of the (reduced) uncertainty bands obtained assuming such correlation.

%%%%%%%%%%%%%%%%%
\begin{figure}[t]
\centering
\includegraphics[width=\rescaletwoplots]{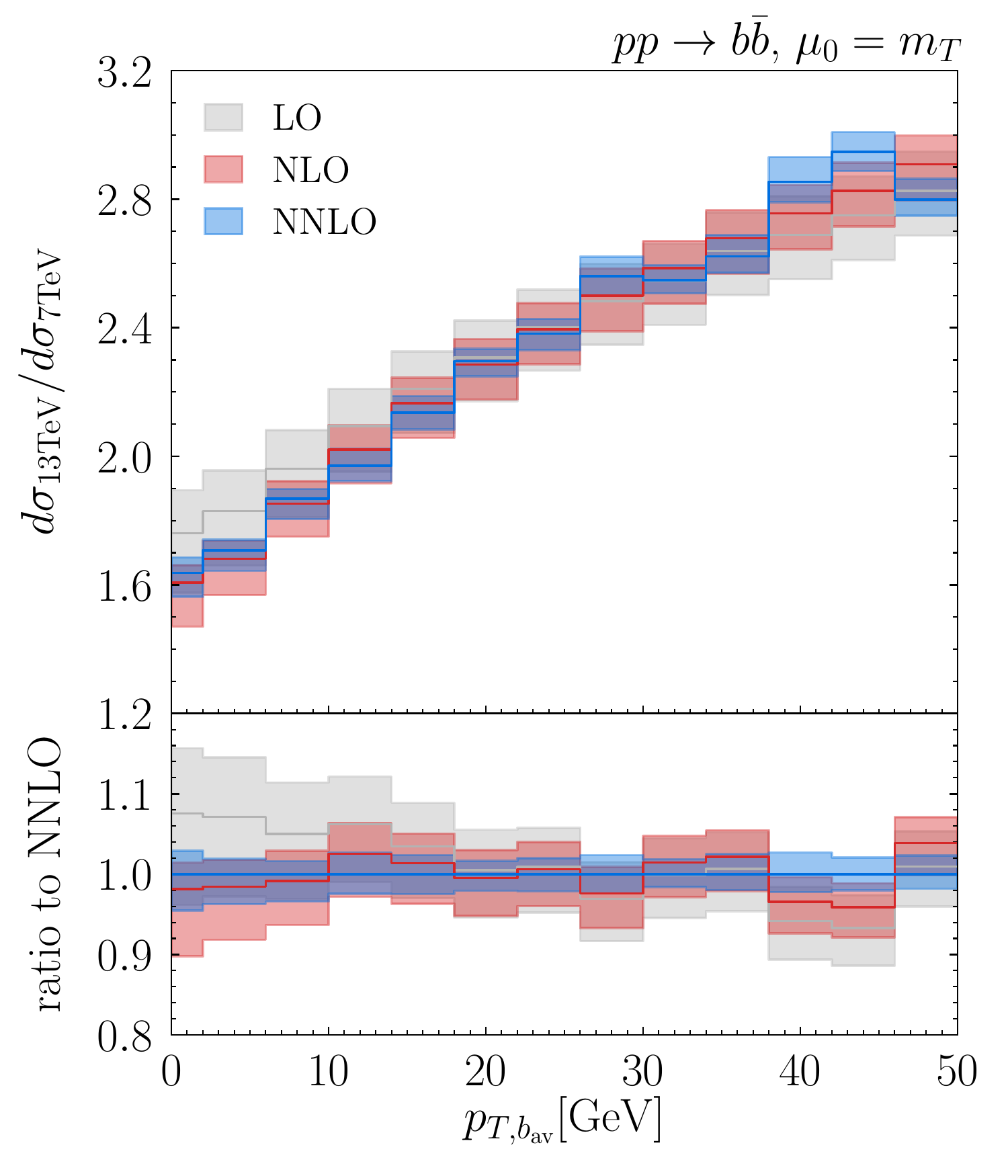}\hspacebetweentwoplots
\includegraphics[width=\rescaletwoplots]{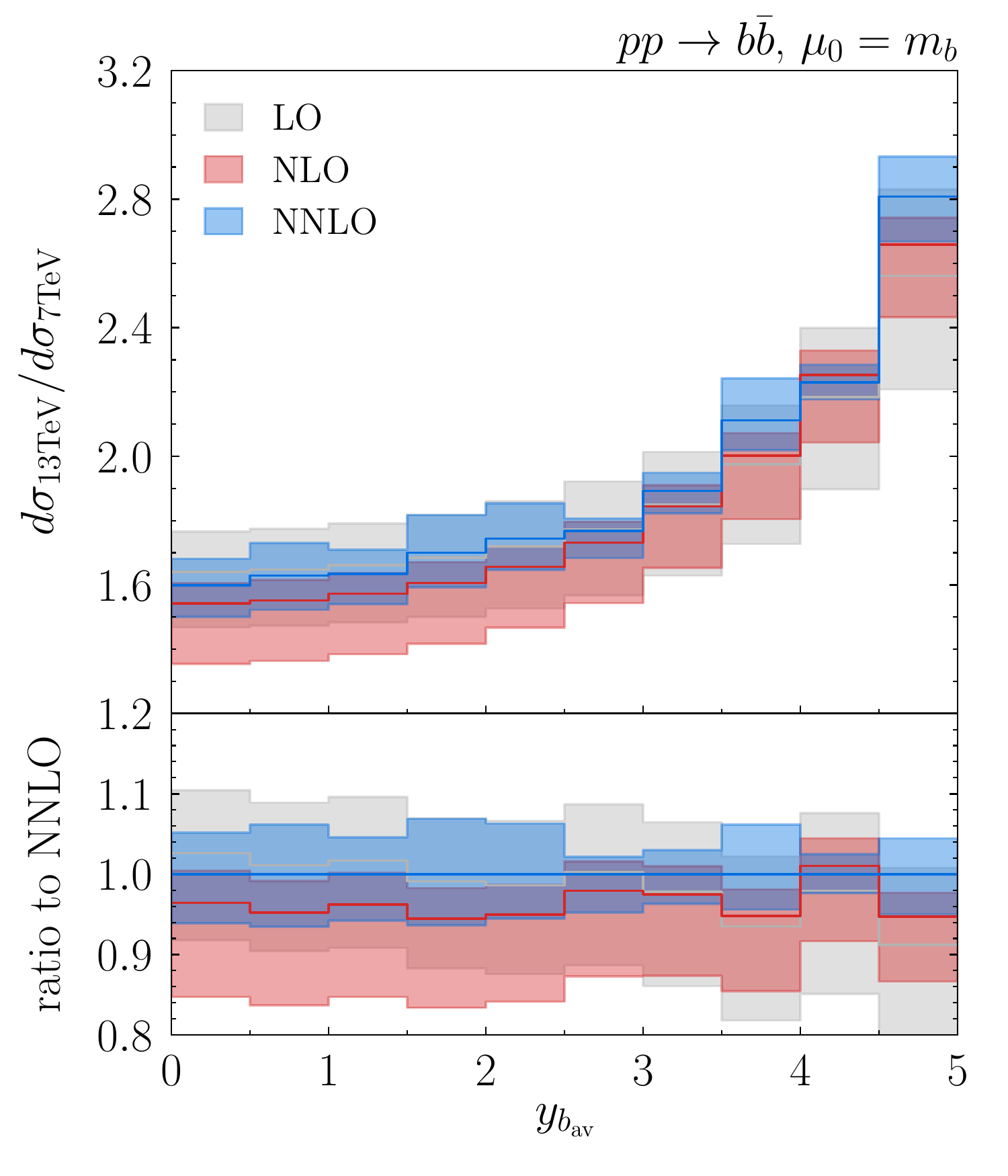}
\spaceabovefigurecaption
\caption{Ratios of 13 TeV to 7 TeV predictions for the transverse momentum (left) and rapidity (right) distributions at the LHC. The lower panels show the result normalised to the NNLO prediction.}
\label{fig:ratio.LHC}
\end{figure}
%%%%%%%%%%%%%%%%%
In Fig.~\ref{fig:ratio.LHC} (left) we show the ratio of transverse-momentum distributions. As the centre-of-mass energy increases, the $p_{T,b_{\mathrm{av}}}$ distribution becomes (slightly) harder, and, therefore, the ratio increases as $p_{T,b_{\mathrm{av}}}$ increases. Considering perturbative uncertainties,
we notice a strong reduction of the uncertainty bands: while the width of the NNLO band in the original distributions ranges from about $\pm 30\%$ at low $p_{T,b_{\mathrm{av}}}$ to about $\pm 10\%$ in the tail, in the ratio the scale uncertainties are reduced to about $\pm 5\%$.
In Fig.~\ref{fig:ratio.LHC} (right) we show the corresponding ratio of rapidity distributions. We see that the ratio increases as the rapidity increases,
consistently with the fact that at 13 TeV the bottom quarks have a stronger tendency to be produced in the forward direction.
Also in this case the LO, NLO and NNLO bands overlap and their size decreases as the order increases, leading to ${\cal O}(\pm 5\%)$ uncertainties at NNLO.
Comparing the results at different perturbative orders for both the transverse-momentum and rapidity distributions, we observe an overlap between the LO, NLO and NNLO uncertainty bands over the full kinematical range. This remarkable stability of the perturbative expansion tends to confirm the approach of computing the ratio by using correlated scale variations at different collider energies.

We close this section with an investigation of bottom-quark production in the forward region.
In Ref.~\cite{Aaij:2016avz}, the LHCb Collaboration has presented results for the measurement of the $b$-hadron production cross section at the LHC.
This measurement used semileptonic decays of $b$-hadrons into a charmed hadron associated with a muon.
The data were collected at $\sqrt{s}=7$~TeV and $13$~TeV, in the pseudorapidity interval $2<\eta<5$, and inclusively in the $b$-hadron transverse momentum.
Measurements of the ratio between the $7$~TeV and $13$~TeV rates were provided as well.

\begin{figure}[t]
\centering
\includegraphics[width=\rescaletwoplots]{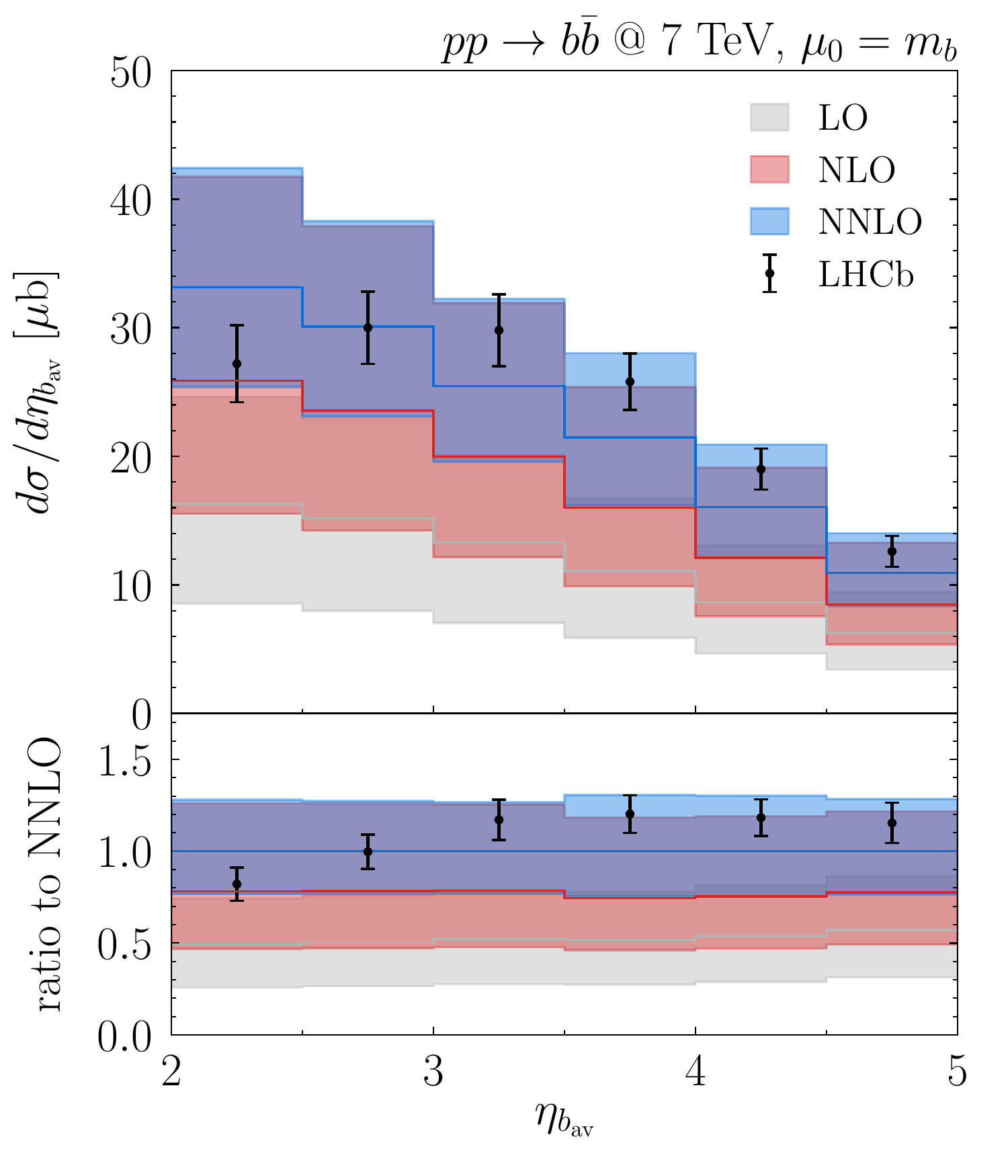}\hspacebetweentwoplots
\includegraphics[width=\rescaletwoplots]{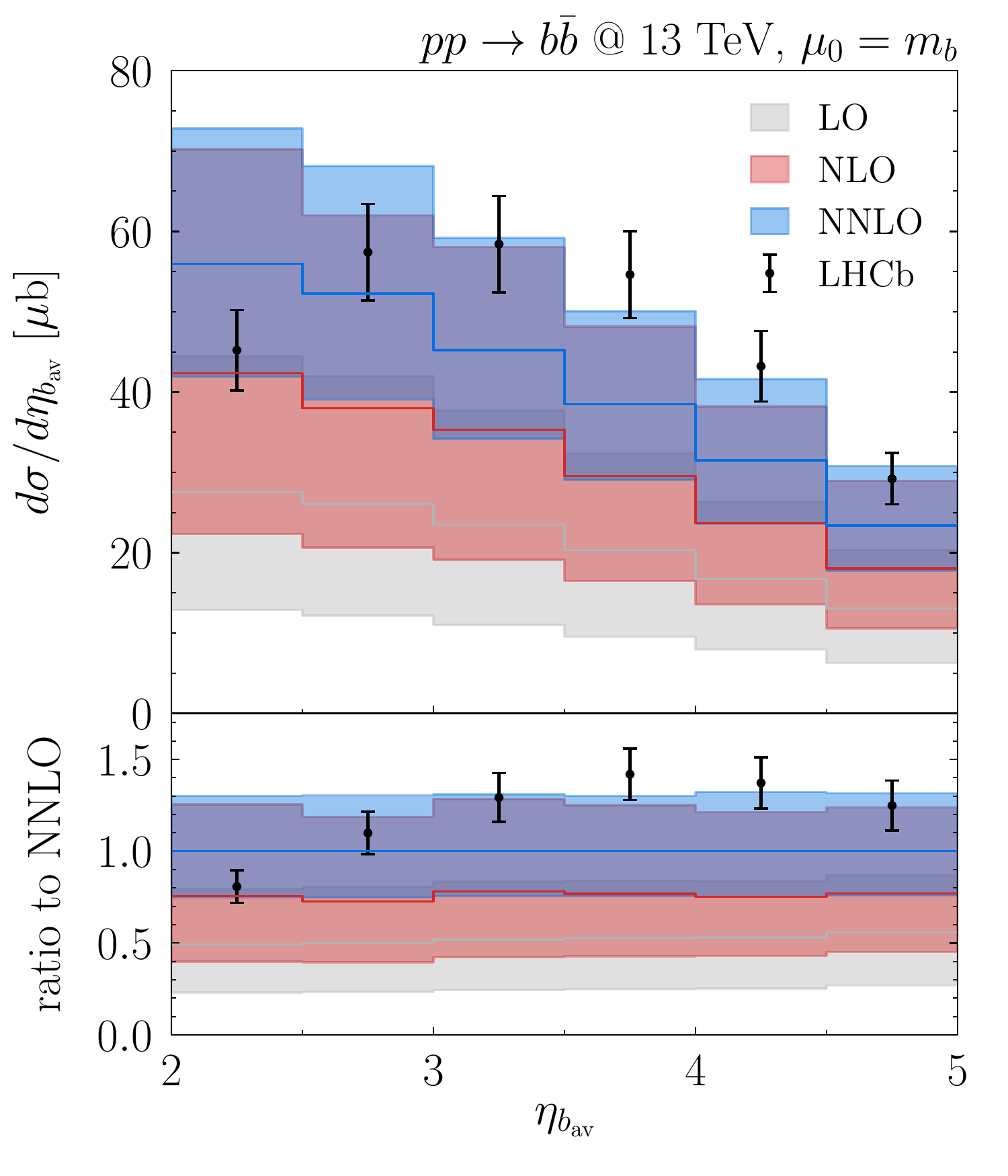}
\spaceabovefigurecaption
\caption{Pseudorapidity distribution at the LHC for the scale choice $\mu_0=m_b$, for centre of mass energy $\sqrt s = 7$~TeV (left) and $13$~TeV (central). The lower panels show the ratio to the NNLO predictions. The theoretical predictions are compared with the LHCb data of Ref.~\cite{Aaij:2016avz}.}
\label{fig:eta_av.LHC}
\end{figure}

Having in mind the limitations of a comparison between theoretical predictions for bottom-quark production and $b$-hadron production data (as already mentioned in Sec.~\ref{sec:resu:tev}),
in Fig.~\ref{fig:eta_av.LHC} we present our perturbative predictions for the pseudorapidity distribution of the bottom quark and
compare them with the experimental data of Ref.~\cite{Aaij:2016avz}.
%%%%%%%%%%%%%%%%%
%%%%%%%%%%%%%%%%%

We first comment on the theoretical results.
As previously noticed, the higher order corrections to the pseudorapidity distributions present similar features to those observed for the rapidity spectrum.
The corrections are almost independent of the value of $\eta_{b_{\rm av}}$ and quantitatively very similar to those
affecting the total cross sections at both collider energies. At NNLO the scale uncertainty is smaller than at NLO, and the corresponding uncertainy bands largely overlap.
The NNLO $K$-factor is reduced with respect to its NLO equivalent.
The effects due to the hadronization of the bottom quarks into $b$-hadrons are expected to be relatively small (see \ref{app:FONLL}),
at the few-percent level in the current range of pseudorapidity and, therefore, well below the perturbative uncertainties at NNLO.

We now comment on the comparison with data. The measured distributions at both collider energies, shown in Fig.~\ref{fig:eta_av.LHC}, are consistent with our NNLO results in the entire pseudorapidity range.
While there is an overlap between NLO prediction and data in almost all cases, the inclusion of the NNLO corrections generally improves the agreement with the central predictions and always strongly reduces the scale uncertainties, making the comparison with data more significant. We note, however, that the
NNLO scale uncertanties are still considerably larger than the experimental errors.

At both collider energies and independently on the perturbative order, we observe that the shapes of the predicted distributions are different from the measured spectra,%
\footnote{Apparently, the data exhibit a maximum at $\eta\sim3$, which is shifted by about one pseudorapidity unit with respect to the theoretical prediction (see Fig.~\ref{fig:neweta_av.LHC}).}
while all individual points are compatible with the data:
at low pseudorapidity values the measurement is below the predicted central value, whereas in the tail the data are closer to the upper bound of the scale uncertainty band. We also observe that the agreement between the NNLO predictions and the data
is slightly worse in the tail of the distribution at $13$~TeV, although
the theory prediction is compatible with the data in each bin.

%%%%%%%%%%%%%%%%%
\begin{figure}[t]
\centering
\includegraphics[width=\rescaletwoplots]{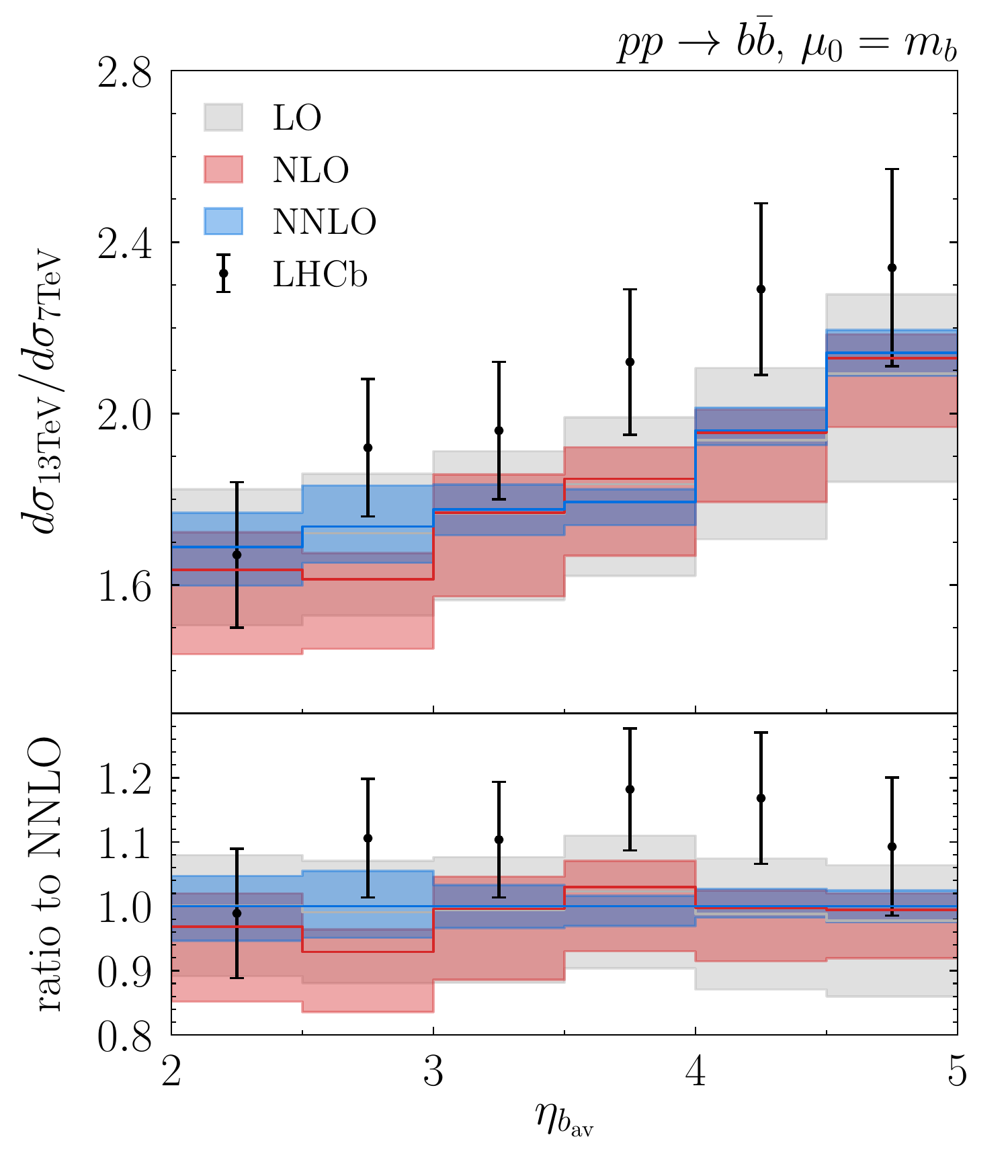}
\spaceabovefigurecaption
\caption{Ratio of 13 TeV to 7 TeV predictions for the pseudorapidity distribution in the forward region. The lower panels show the result normalised to the NNLO prediction. The theoretical predictions are compared with the LHCb data of Ref.~\cite{Aaij:2016avz}.}
\label{fig:ratio_eta}
\end{figure}
%%%%%%%%%%%%%%%%%
In Fig.~\ref{fig:ratio_eta} we show the ratio of the 13 TeV and 7 TeV predictions for the pseudorapidity distribution, computed as for the results in Fig.~\ref{fig:ratio.LHC}.
As expected, the ratio features a strong reduction of the scale uncertainties at NNLO, from about $\pm 30\%$ to only about $\pm 5\%$. Moreover, the
LO, NLO and NNLO uncertainty bands overlap, thereby supporting again the use of correlated variations of the renormalisation and factorisation scales.
Considering the data--theory comparison, we observe that
the reduced NNLO scale uncertainties of the ratio predictions are significantly smaller than the experimental errors.

Comparing the NNLO prediction with the experimental data, we observe that, with the exception of the first bin, the data are systematically above the central NNLO result. In two bins the data points lie outside of the NLO and NNLO uncertainty bands.
We note, however, that the $\eta$-independent systematic errors are the dominant source of uncertainty in the measurement of the ratio: while the total (statistical plus systematic) uncertainty ranges between $\pm 8.2\%$ and $\pm 10.5\%$ for the different bins, the $\eta$-independent uncertainty is $\pm 7.4\%$ (see Tables~3 and 4 of Ref.~\cite{Aaij:2016avz}).
This $\eta$-independent uncertainty is of the same order as the observed data--theory discrepancy.
In the data--theory comparison of Fig.~\ref{fig:eta_av.LHC} we had observed similar shape differences at $7$~TeV and $13$~TeV, which largely cancel out in the ratio.

The pseudorapidity distribution can be integrated over the range $2<\eta_{b_{\rm av}}<5$ to obtain the accepted cross section.
\begin{table}[tb]
\begin{center}
\renewcommand{\arraystretch}{1.6}
\setlength{\tabcolsep}{1em}
\begin{tabular}{|c|l|l|l|}
\hline
\multicolumn{1}{|c|}{$\sigma\;[\mu\text{b}]$} & \multicolumn{1}{c|}{$pp$ @ $7$ TeV} & \multicolumn{1}{c|}{$pp$ @ $13$ TeV}&\multicolumn{1}{c|}{Ratio}\\ \hline
NNLO ($\mu_0=m_b$)  & $68.5(4)\phantom{.}^{+28\%}_{-24\%}$ & $123.4(7)\phantom{.}^{+30\%}_{-25\%}$ & $1.80(2)\phantom{.}^{+2.4\%}_{-3.6\%}$ \\
NNLO ($\mu_0=2m_b$) & $61.9(2)\phantom{.}^{+30\%}_{-18\%}$ & $111.6(4)\phantom{.}^{+18\%}_{-17\%}$& $1.80(1)\phantom{.}^{+1.1\%}_{-1.6\%}$ \\ \hline
&&&\\[-4ex]
LHCb         &
$72.0\begin{array}{l}\\[-5.2ex] \pm 0.3\;({\rm stat}) \\[-1.6ex] \pm 6.8\;({\rm syst})\end{array}$ &
$144\begin{array}{l}\\[-5.2ex] \pm \phantom{0}1\;({\rm stat}) \\[-1.6ex] \pm 21\;({\rm syst})\end{array}$ &
$2.00\begin{array}{l}\\[-5.2ex] \pm 0.02\;({\rm stat}) \\[-1.6ex] \pm 0.26\;({\rm syst})\end{array}$ \\ \hline
\end{tabular}
\end{center}
\vspace*{-0.4cm}
\caption{Cross section for single-inclusive $b$-quark production at $\sqrt{s}=7$ and 13 TeV in the pseudorapidity
region $2<\eta_{b_{\rm av}}<5$. The NNLO results are compared with the LHCb measurement of
Ref.~\cite{Aaij:2016avz}. Numerical errors on the last digits are stated in brackets,
as in Table~\ref{table:total_vs_hathor}.
The LHCb data are presented with their statistical and systematic uncertainties.
}
\label{table:eta.LHC}
\end{table}

In Table \ref{table:eta.LHC} the corresponding NNLO predictions at $\sqrt{s}=7$ and 13 TeV
as well as their ratio are compared with the LHCb data. The NNLO result is stated for two values of central scales, $\mu_0=m_b$ and $\mu_0=2m_b$, and we find the same ratio of the 13 and 7 TeV predictions
for these two scale choices. The result for $\mu_0=m_b$ suggests larger (and thus more conservative) perturbative uncertainties.
Comparing the NNLO predictions for the integrated cross section to the LHCb measurement, we find that both scale choices lead to predictions that are consistent with the data,
the choice $\mu_0=m_b$ leads to a better agreement though. The NNLO prediction for the ratio is in good agreement with the experimental measurement.

Considering the $b{\bar b}$ total cross section at NNLO,
in Table~\ref{table:totalXSwithunc} we have reported its PDF uncertainty
($\Delta_{\rm PDFs}$) and scale variation uncertainty
($\Delta_{\rm scale}$), showing that $\Delta_{\rm PDFs}$ is definitely smaller than $\Delta_{\rm scale}$. In this paper we do not present a study of PDF uncertainties
on differential cross sections. We note that, in the very-forward region,
PDF uncertainties on (pseudo)rapidity differential cross sections (and on their ratio at different energies) can be larger than the corresponding uncertainties on the total cross section, and their size can be similar to the size of the scale variation uncertainties that we find at NNLO.
We remark on this fact by quoting, for instance, some values of $\Delta_{\rm PDFs}$
from the NLO study of Ref.~\cite{Cacciari:2015fta}. The PDF uncertainties tend to slightly increase as $\sqrt s$ increases (see accompanying comments to Table~\ref{table:totalXSwithunc}). In the case of $d\sigma/d\eta$ at ${\sqrt s}=13$~TeV, the
NLO value of $\Delta_{\rm PDFs}$ is about $\pm 9\%$ at $\eta \sim 4$ and increases to about $\pm 18\%$ at $\eta \sim 5$~\cite{Cacciari:2015fta}
(we recall that we find $\Delta_{\rm scale} \sim \pm 30\%$ at NNLO, see
Fig.~\ref{fig:eta_av.LHC}). In the case of the ratio of $d\sigma/d\eta$
at 13~TeV and 7~TeV, the NLO value of $\Delta_{\rm PDFs}$ is of ${\cal O}(1\%)$
at $\eta \ltap 3$ and increases to about $\pm 5\%$ ($\pm 8\%$) at $\eta \sim 4$
($\eta \sim 5$)~\cite{Cacciari:2015fta} (at NNLO we find that $\Delta_{\rm scale}$
varies in the range between $\pm 5\%$ and $\pm 3\%$, see Fig.~\ref{fig:ratio_eta}).

%=============================================================================================

%=============================================================================================
\section{Summary}
\label{sec:summa}
In this paper we have presented the first fully differential NNLO calculation of bottom-quark pair production at hadron colliders.
The calculation was carried out by using the $q_T$ subtraction formalism to handle and cancel IR divergences. It extends the corresponding calculation for top-quark pair production~\cite{Catani:2019iny,Catani:2019hip}.
The computation has been implemented in the {\sc Matrix} framework, which allows us to evaluate single- and multi-differential distributions with arbitrary IR safe selection cuts.

We have presented results for the total cross section at the Tevatron and the LHC and compared them with predictions obtained by using the numerical code {\sc Hathor}, finding excellent agreement.
We have studied different sources of theoretical uncertainty and observed that, despite the inclusion of NNLO corrections,
perturbative uncertainties are still sizeable and dominant over other sources of uncertainties.

We have presented predictions for single-differential distributions at the Tevatron ($\sqrt{s}= 1.96$ TeV) and at LHC ($\sqrt{s} = 7$ TeV and $\sqrt{s} = 13$ TeV).
As a general feature, we observed that the inclusion of NNLO corrections suggests a (slow) convergence of the perturbative series, with a good overlap between NLO and NNLO bands and a significant reduction of perturbative uncertainties, which are estimated as usually through scale variations.
Since perturbative uncertainties at NNLO are still large also for differential distributions, we have investigated possible ways to reduce them.
We considered both normalised distributions and ratios of distributions at different energies, and we assumed scale uncertainties to be fully correlated in the respective ratios.
In both cases we showed that the ensuing results are perturbatively stable and that LO, NLO and NNLO uncertainty bands overlap, suggesting that this approach indeed provides reliable predictions with reduced perturbative uncertainties.

We have also compared our predictions with experimental measurements for $b$-hadron production from the CDF Collaboration at the Tevatron and the LHCb Collaboration at the LHC, finding reasonably good agreement.
Further studies in the high-$p_T$ region require the resummation of large logarithmic contributions of the form $\ln p_T/m_b$, while more detailed data--theory comparisons could benefit from the inclusion of fragmentation effects.
Such studies are left for future work.
%=============================================================================================

%=============================================================================================
\noindent {\bf Acknowledgements}

\noindent
We thank Matteo Cacciari, Katharina M{\"u}ller and Giovanni Passaleva for useful discussions and comments on the manuscript.
We are indebted to Federico Buccioni, Jean-Nicolas Lang, Jonas Lindert and Stefano Pozzorini for their ongoing support with {\sc OpenLoops~2}, and in particular for making specific amplitudes available to us.
We are also grateful to Emanuele Nocera and the NNPDF collaboration for providing us with the PDF grids needed to evaluate $\as$ uncertainties.
This work is supported in part by the Swiss National Science Foundation (SNF) under contracts 200020\_188464 and IZSAZ2\_173357.
The work of SK is supported by the ERC Starting Grant 714788 REINVENT.
%=============================================================================================

\appendix
\gdef\thesection{Appendix \Alph{section}}

%=============================================================================================
\section{Shape of the pseudorapidity distribution}
\label{app:eta-y}
In this Appendix we briefly recall the origin of shape differences between
rapidity and pseudorapidity distributions for massive particles.

We consider the inclusive production of a single particle of mass $m$ and transverse momentum $p_T$, and we relate its rapidity $y$ and pseudorapidity $\eta$, as defined in the centre--of--mass frame of the two colliding particles in the initial-state. If
$m=0$, we have $\eta=y$. If the produced particle has a non-vanishing mass, its pseudorapidity $\eta=\eta(y,p_T/m)$ and rapidity $y$ can be directly related at {\it fixed} values of $p_T$, and this relation is controlled by the parameter $p_T/m$.
We explicitly have
\begin{equation}\label{eq:etavsy}
\eta(y,p_T/m) = \ln \left[ \frac{m_T}{p_T} \sinh y
+ \sqrt{ 1 + \left(\frac{m_T}{p_T}\right)^2 \sinh^2 y } \;
\right] \;\;,
\end{equation}
where $m_T$ is the transverse mass ($m_T^2=m^2+p_T^2$).

The relation (\ref{eq:etavsy}) is symmetric under inversion ($\eta \leftrightarrow -\eta, \, y \leftrightarrow -y$) and we limit ourselves to comment on the region of positive values of both $\eta$ and $y$.
If $y=0$ we have $\eta=0$. Increasing the value of $y$, the difference $\eta -y$ monotonically increases and reaches its maximal value $\eta - y \simeq \ln(m_T/p_T)$
at $y \gg 1$. Therefore, events at finite values of $y$ are moved to larger values of $\eta$ and the shift $\eta(y,p_T/m) - y$ increases by either increasing $y$ or decreasing $p_T$. This kinematical effect produces shape variations between the rapidity and pseudorapidity distributions.

The rapidity and pseudorapidity cross sections at the double-differential level (i.e., at fixed $p_T$) are related as follows,
\begin{equation}\label{eq:doublediff}
\frac{d\sigma}{dp_T \,d\eta} = J(y,p_T/m) \; \frac{d\sigma}{dp_T \,dy} \;,
\end{equation}
where the Jacobian function $J(y,p_T/m)$ is
\begin{equation}\label{eq:jac}
J(y,p_T/m) = \left( \frac{d\eta}{dy} \right)^{-1} =
\sqrt{1-\frac{m^2}{m_T^2 \cosh^2y}} \;.
\end{equation}
Note that the function $J(y,p_T/m)$ monotonically increases as $|y|$ increases, and it varies in the range $(p_T/m_T) \leq J < 1$. Therefore, the $y$ dependence of the Jacobian factor in the right-hand side of Eq.~(\ref{eq:doublediff}) is qualitatively opposite to the $y$ dependence of the rapidity cross section ($d\sigma/(dp_T \,dy)$
monotonically decreases as $|y|$ increases), and this produces a maximum of the expression in  Eq.~(\ref{eq:doublediff}) at a non-vanishing value of $|y|$ and, hence, of $|\eta|$. Moreover, at $\eta=y=0$, the pseudorapidity and rapidity cross sections are related as
\begin{equation}\label{eq:doubledip}
\left( \frac{d\sigma}{dp_T \,d\eta}\right)_{\eta=0} = r_0(p_T/m) \;
 \left( \frac{d\sigma}{dp_T \,dy} \right)_{\eta=0} \;,
\quad \quad \quad r_0(p_T/m) = \frac{p_T}{m_T} \;,
\end{equation}
and the factor $r_0(p_T/m) < 1$ leads to the central-pseudorapidity dip of
$d\sigma/(dp_T \,d\eta)$.

The differences between the double-differential pseudorapidity and rapidity distributions have a quantitative dependence on the actual value of $p_T$ (and $m$),
but such dependence is `monotonic': the differences are maximal if $p_T \ll m$, and they tend to vanish if $p_T \gg m$
(since $\eta \simeq y$ in this region). Therefore,
the main features that we have just discussed (shifted maximun and central dip of the pseudorapidity cross section) remain valid at the single-differential level after integration over $p_T$.

Such features are clearly visible for bottom-quark production by comparing the
single-differential cross sections
$d\sigma/dy$ and $d\sigma/d\eta$ in Figs.~\ref{fig:y_av.LHC}
and~\ref{fig:neweta_av.LHC}, respectively. The pseudorapidity distribution has its maximum value at $\eta\sim 1.5$--$2.0$, and at central pseudorapidity we have
\begin{equation}\label{eq:singledip}
\left( \frac{d\sigma}{d\eta}\right)_{\eta=0} = {\overline r}_0 \;
 \left( \frac{d\sigma}{dy} \right)_{\eta=0} \;,
\end{equation}
with the value ${\overline r}_0 \simeq 0.7$ (at both NLO and NNLO in the bin where
$0 \leq \eta \leq 0.5$) that is significantly smaller than unity.
In the case of bottom-quark production the $p_T$ differential cross sections are highly dominated by their values in the region where $p_{T, b/{\bar b}}$ is of the order of $m_b$ (see Figs.~\ref{fig:pT_av.Tev}--\ref{fig:pT_av.LHC}). Therefore,
relations between single-differential distributions can be roughly obtained
from Eqs.~(\ref{eq:doublediff}) and (\ref{eq:doubledip}) by performing the
replacements $d\sigma/(dp_T \, dx) \to d\sigma/dx$ (with $x=\eta,y$) and $J \to
\langle J \rangle$, where $\langle J \rangle$ is the `average' (typical) value of $J$
that is obtained by the $p_{T, b/{\bar b}}$ integration over the region with
$p_{T, b/{\bar b}} \sim m_b$.
In particular, in Eq.~(\ref{eq:doubledip}) this replacement leads to
Eq.~(\ref{eq:singledip}) with
${\overline r}_0 = \langle p_{T, b/{\bar b}}/m_{T, b/{\bar b}} \rangle_0$,
where $\langle p_T/m_T \rangle_0$ is the average value of $p_T/m_T$ with respect to
the $p_T$ integration of $d\sigma/(dp_T \,dy)$ at $y=0$. Such expression is consistent
with the numerical value ${\overline r}_0 \simeq 0.7$
(we recall that the average transverse momenta of the $b$ quark are 5.5~GeV
and 5.9~GeV at the two LHC energies that we are considering) that we find
in Eq.~(\ref{eq:singledip})
from the results in Figs.~\ref{fig:y_av.LHC} and~\ref{fig:neweta_av.LHC}.

We remark on the fact that the relation in Eq.~(\ref{eq:doubledip}) between
double-differential distributions has an entirely kinematical origin. However, the relation between the single-differential cross sections $d\sigma/dy$ and
$d\sigma/d\eta$ (e.g., Eq.~(\ref{eq:singledip})) also includes some dynamical effects, since it is controlled by the typical value of $p_T/m$, which does depend on the production dynamics of the observed particle.

To qualitatively highlight these dynamical effects we can compare, e.g.,
bottom-quark and pion ($\pi$) production. In both cases, $d\sigma/d\eta$ has a maximum at a finite value of $|\eta|$ and a central-pseudorapidity dip. However, the shape differences between $d\sigma/dy$ and $d\sigma/d\eta$ are much less pronounced in the case of pion production since $\langle p_{T, \pi} \rangle/m_\pi$ is sizeably larger than $\langle p_{T, b/{\bar b}} \rangle/m_b$.
At LHC energies we have $\langle p_{T, \pi}\rangle \sim 400~{\rm MeV} \sim 3 m_\pi$
and (consequently) ${\overline r}_0 \sim 0.95$ in Eq.~(\ref{eq:singledip}).
%=============================================================================================

%=============================================================================================
\section{Comparison with FONLL calculations}
\label{app:FONLL}
The so called FONLL prediction \cite{Cacciari:1998it,Cacciari:2001td} has been employed in a variety of data--theory comparisons for bottom (and charm) production at the Tevatron and the LHC.
At large transverse momenta of the heavy quark $Q$ (i.e. $p_T\gg m_Q$) large logarithmic contributions due to
multiple quasi-collinear emissions appear in the perturbative calculation.
Such terms, which have the form $\as^n \ln^k (m_T/m_Q)$ ($k\leq n$), need to be resummed to all orders.
The FONLL prediction is obtained by resumming them
up to next-to-leading logarithmic accuracy, and the resummed result is combined and consistently matched to the
NLO result
to avoid double-counting of perturbative contributions up to ${\cal O}(\as^3)$.
The resummation of the large logarithmic terms is carried out by folding the partonic cross section for the inclusive production of a high-$p_T$ massless parton $i$
(the heavy quark is also considered to be effectively massless at high $p_T$)
with the perturbative function $D_{i\to Q}(z,\mu_F)$ \cite{Mele:1990cw}
that describes the fragmentation of the massless parton at scale $\mu_F$ into the massive heavy quark $Q$ at a scale of the order of $m_Q$. The perturbative fragmentation function depends on the factorization scale $\mu_F$ and on the momentum fraction $z$ that is transferred to the produced heavy quark $Q$.

In the following, we present a comparison of our NLO and NNLO results with the FONLL result, which is obtained from the code in Ref.~\cite{webpage}.
For this comparison we limit ourselves to considering $b$-quark production at the LHC with $\sqrt s=7$~TeV.
The FONLL results are obtained by using the NNPDF30 NLO PDFs \cite{Ball:2014uwa}
with $n_f=5$ massless-quark flavours and $\as(m_Z)=0.118$.
The value of $m_b$ is fixed to $m_b=4.75$ GeV, and the central scale is $\mu_0=m_T$
for both the renormalisation and factorisation scales.
Our NLO and NNLO results are obtained with the same parameters by using the NNPDF30 NLO and NNLO PDFs with $n_f=4$ massless-quark flavours.%
\footnote{We note that since the NLO calculation entering the FONLL prediction is carried out with $n_f=5$ PDFs the two NLO results do not exactly coincide.
By comparing our NLO results with those obtained from Ref.~\cite{webpage} we indeed find differences ranging from the $1-2\%$ level at low $p_T$ to the $5\%$ level at $p_T=50$ GeV.}

%% total cross section

We find that the $p_T$ integral of $d\sigma/dp_T$ increases by about 7\% in going from NLO to FONLL accuracy. We note that this increase of the total cross section\footnote{Precisely speaking, the integral of the $p_T$ differential cross section gives the total cross section times the value of the average multiplicity  of $b$-quarks.}
is not a significant prediction of the FONLL calculation (since the total cross section
does not have collinear logarithmic contributions to be resummed), and it has to be regarded as a (higher-order) systematic effect due to the arbitrariness of the procedure that is used to combine the NLO and resummed calculations in the low-$p_T$ region. We also note that such systematic effect is `acceptable', since its size is definitely smaller than the size of the NLO perturbative uncertainty from scale variations (see Table~\ref{table:totalXS}).

%%%%%%%%%%%%%%%%%%%%
\begin{figure}[t]
\centering
\includegraphics[width=\rescaletwoplots]{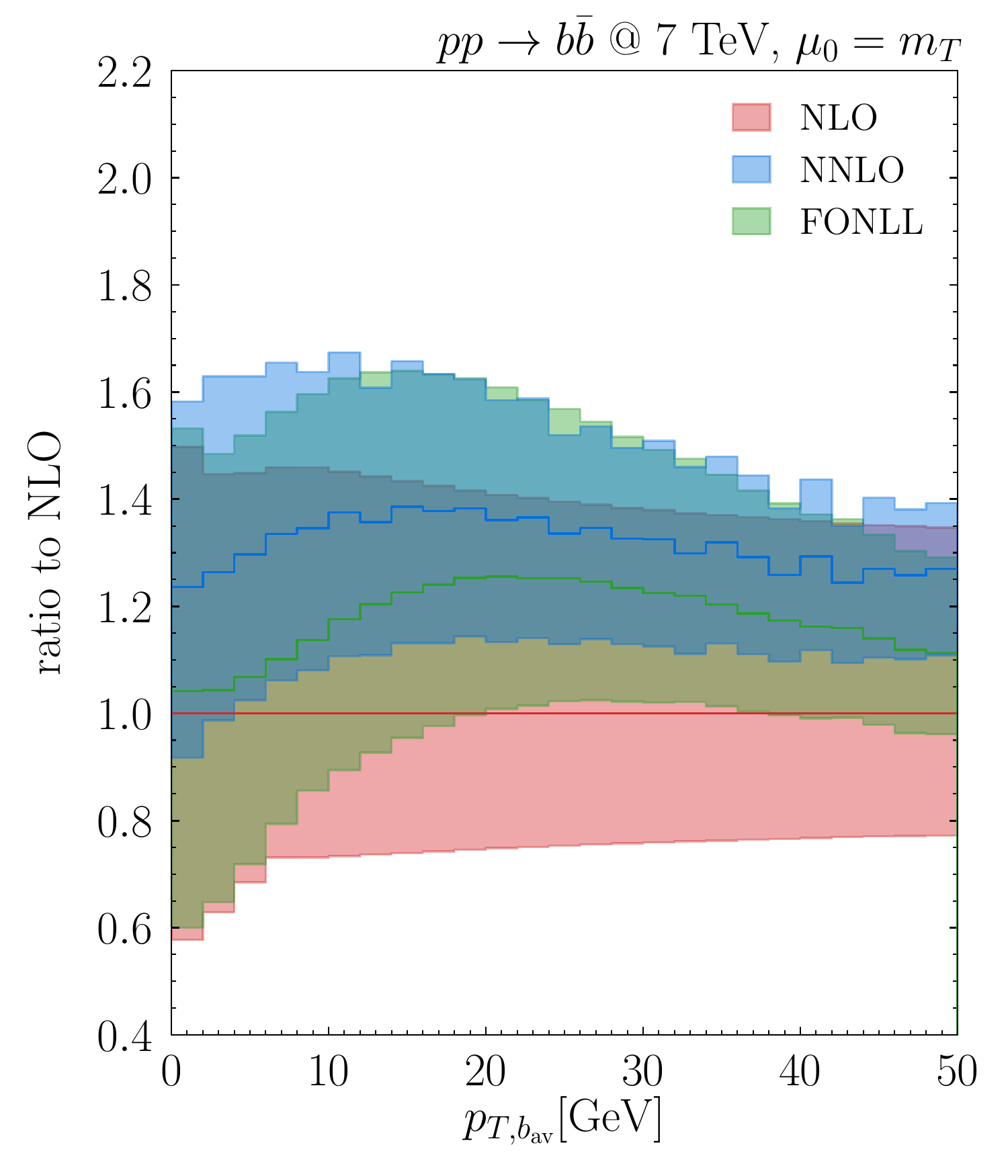}
\spaceabovefigurecaption
\caption{Comparison between NLO, FONLL and NNLO predictions for the transverse-momentum distribution $d\sigma/dp_T$ of the bottom quark at the LHC (${\sqrt s}=7$~TeV). The three results with their scale variations are normalised to the NLO prediction at central scale $\mu_0=m_T$.}
\label{fig:fonll-pt}
\end{figure}

The comparison of the $p_T$ distributions at different perturbative orders is shown in Fig.~\ref{fig:fonll-pt}.
We start our discussion by comparing the FONLL and NLO results.
The high-$p_T$ resummation has a non-trivial effect on the shape of the perturbative NLO result.
At small transverse momenta ($p_T \ltap 10$~GeV) the resummation effect is at the few-percent level (this is the $p_T$ region that mostly contributes to the value of the total cross section), and it increases to
about $25\%$ at $p_T\sim 20$--$30$~GeV, decreasing again as $p_T$ increases,
and it is about $10\%$ at $p_T\sim 50$~GeV.
The region of higher values of $p_T$ (which is not shown in the figure) is the most relevant for resummation: here
the effect of resummation becomes negative \cite{Cacciari:1998it} and the FONLL prediction eventually undershoots the NLO result.
This behaviour is due to the fact that multiple radiation of collinear partons tends to make the $p_T$ spectrum softer at high values of $p_T$.
in the low-$p_T$ region ($p_T\ltap 20$ GeV) the FONLL and NLO uncertainty bands are of comparable size, and the effect of resummation is smaller than the scale uncertainties.
In this region, the FONLL and NLO predictions are fully consistent, and their differences are due to higher-order effects that are not predicted by resummation.
At higher values of $p_T$ (30\,GeV\,$\ltap p_T \ltap 50$\,GeV) the main effect of the resummation is a reduction of the scale uncertainties.

We now comment on the comparison between the FONLL and NNLO results. At small transverse momenta the NNLO $K$-factor increases from about 1.2 to about 1.35 at $p_T\sim 15$ GeV and then slightly decreases to about 1.3 at $p_T\sim 50$ GeV.
Over the entire region of transverse momenta shown in Fig.~\ref{fig:fonll-pt}, the NNLO central result is systematically higher than the FONLL result and has a smaller uncertainty, especially in the low- and intermediate-$p_T$ regions.
The FONLL and NNLO uncertainty bands overlap.
Owing to these features, we conclude that the NNLO result is a reliable prediction in the entire region of transverse momenta considered in Fig.~\ref{fig:fonll-pt}. For higher values of $p_T$,
the impact of resummation will start to be important to reduce the perturbative uncertainties. Eventually, for very high tranverse momenta, a resummed calculation will be needed to obtain reliable predictions.

\begin{figure}[t]
\centering
\includegraphics[width=\rescaletwoplots]{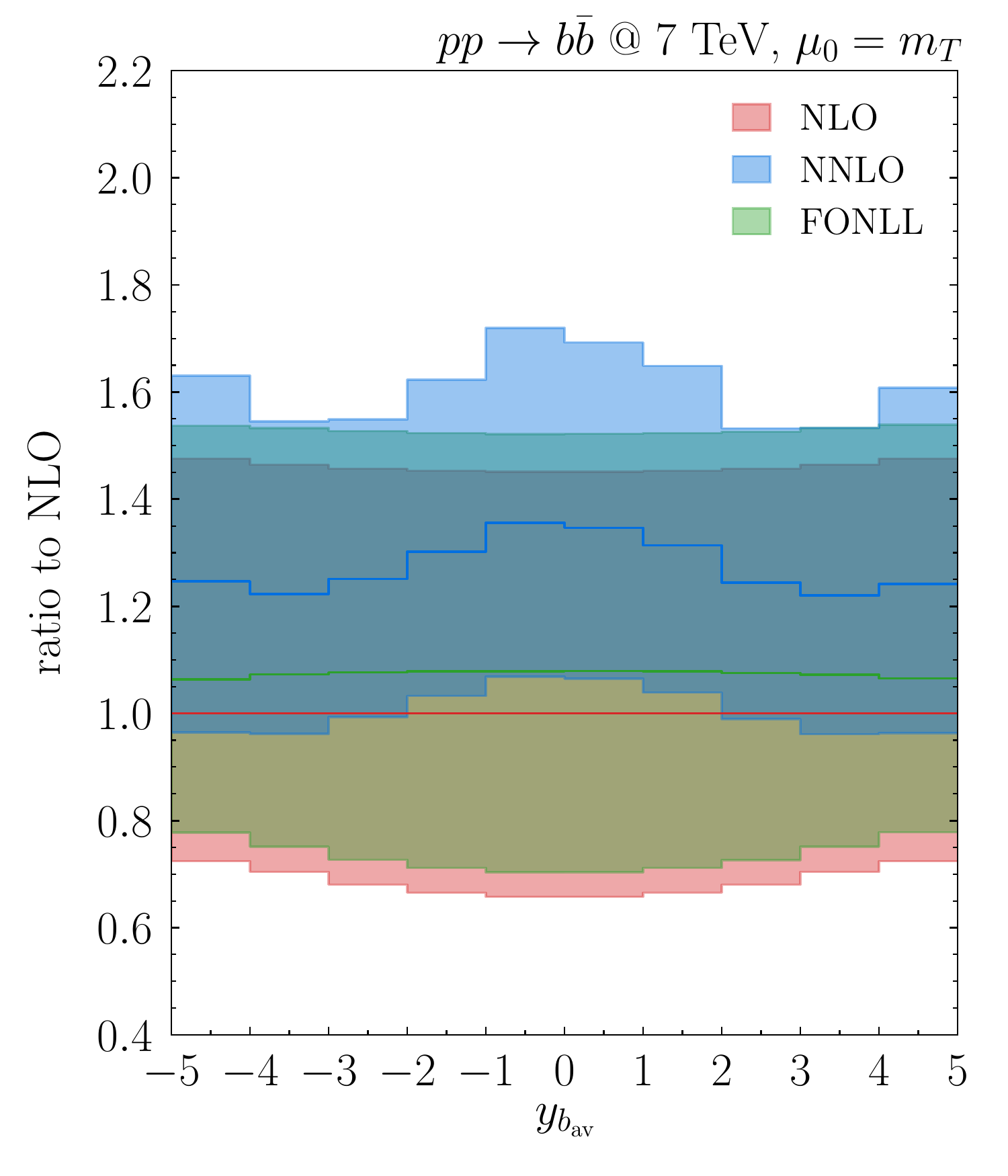}\hspacebetweentwoplots
\includegraphics[width=\rescaletwoplots]{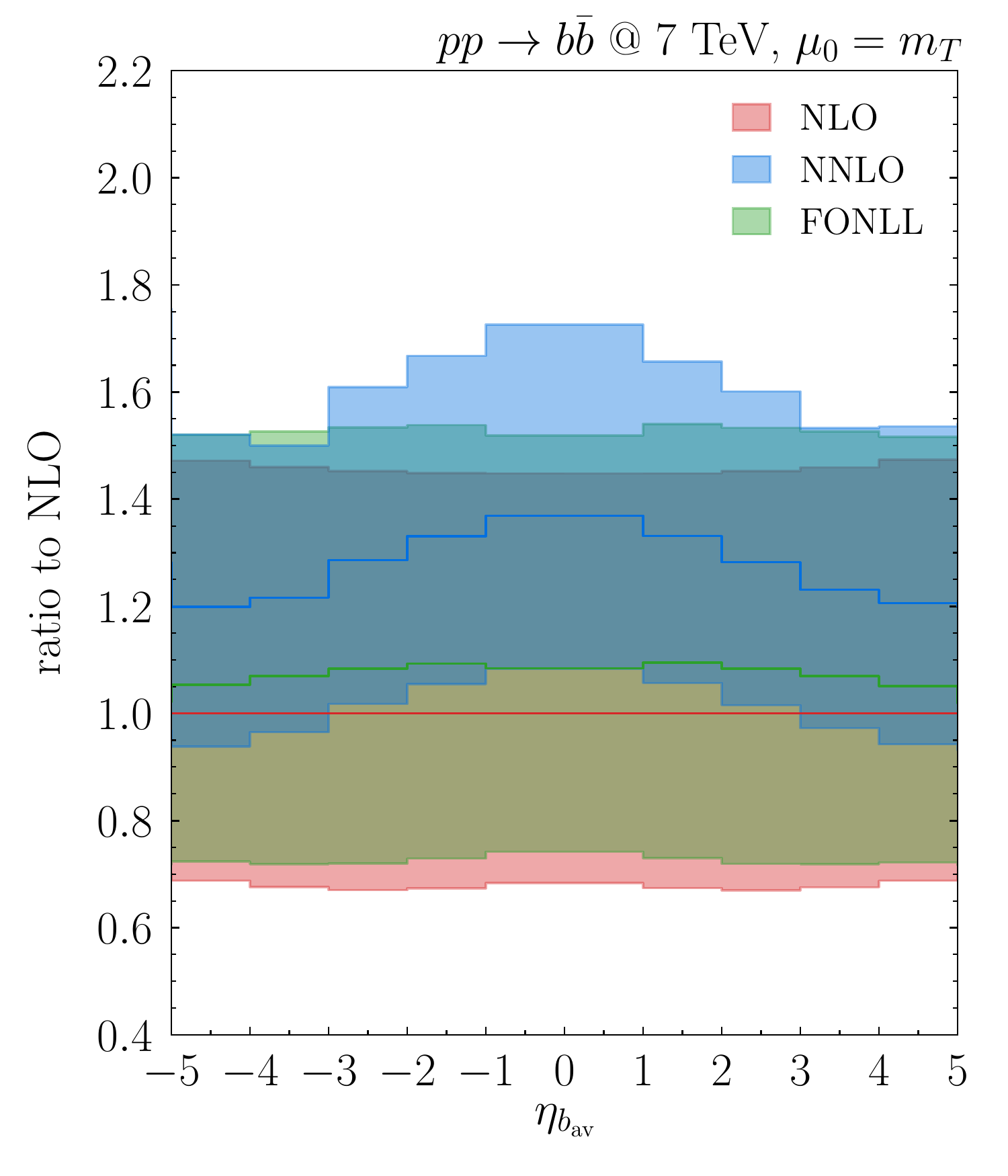}
\spaceabovefigurecaption
\caption{Comparison between NLO, FONLL and NNLO predictions for the rapidity distribution $d\sigma/dy_{b_{\rm av}}$ (left) and pseudorapidity distribution
  $d\sigma/d\eta_{b_{\rm av}}$ (right) of the bottom quark at the LHC (${\sqrt s}=7$~TeV). The three results with their scale variations are normalised to the NLO prediction at central scale $\mu_0=m_T$.}
\label{fig:fonll-etay}
\end{figure}

Comparing the FONLL and NLO
results for the rapidity and pseudorapidity distributions
(Fig.~\ref{fig:fonll-etay}),
we observe that the FONLL result is systematically higher than the NLO result.
Such enhancement is consistent with the increase of the corresponding total cross sections.
For the rapidity distribution the FONLL/NLO ratio is rather flat and about 1.07,
while for the pseudorapidity distribution,
the FONLL/NLO ratio is about 1.08
in the central region and decreases as $|\eta_{\rm av}|$ increases.
In particular, in the region relevant for LHCb data (see Figs.~\ref{fig:eta_av.LHC} and \ref{fig:ratio_eta}),
the FONLL result is about $6\%$ ($3\%$) higher than the NLO result at $\eta_{\rm av}=2$ ($\eta_{\rm av}=5$).
Comparing the scale uncertainty bands at NLO and FONLL accuracies, we see that they have a very similar size, which is definitely larger than the difference between the NLO and FONLL results at central scales.
The relatively small and uniform impact of the resummation on the (pseudo)rapidity distribution is not unexpected.
The perturbative resummation implemented in the FONLL calculation deals with
$\ln (m_T/m_b)$ terms, and thus it should not significantly affect the shape of such  distributions.
The FONLL relative effect of ${\cal O}(10\%)$ with respect to NLO is due to the lower-$p_T$ region in Fig.~\ref{fig:fonll-pt} where resummation cannot improve the predictivity of fixed-order calculations.
In Fig.~\ref{fig:fonll-etay} the NNLO $K$-factor ranges between 1.2 and 1.3, and the NNLO effects are larger than the FONLL effects by ${\cal O}(20\%)$.
The scale uncertainty bands of the FONLL and NNLO results do overlap, and the NNLO result has smaller uncertainties over the entire (pseudo)rapidity region.

% non-perturbative fragmentation

The fixed-order and FONLL results that we have so far discussed in this Appendix refer to $b$-quark production.
Within the FONLL framework, cross sections for the inclusive production of a single $B$ hadron can be computed by supplementing the corresponding $b$-quark cross section with non-perturbative effects that describe the fragmentation of the $b$ quark into the $B$ hadron. These effects are implemented in the double-differential cross section with respect to the transverse momentum $p_T$ and rapidity $y$ of the produced particle
(either $b$ quark or $B$ hadron) by performing a convolution of the $b$-quark cross section with
the non-perturbative fragmentation function of the $b$ quark.
The convolution acts on the fraction of the transverse momentum that is transferred
from the $b$ quark to the $B$ hadron, while the rapidity is kept fixed (i.e., the rapidities of the $b$ quark and $B$ hadron are set to be equal).
The multiplicity of $B$ hadrons from the non-perturbative fragmentation function
is set to be equal to unity. Using this procedure, the non-perturbative fragmentation
produces no effects on the single-differential cross section $d\sigma/dy$ and, consequently, on the total cross section. Non-perturbative fragmentation effects
are instead introduced in other differential distributions.

Using the FONLL code of Ref.~\cite{webpage}, we have computed the effects of
non-perturbative fragmentation on the single-differential cross sections with respect to the transverse momentum and to the pseudorapidity of the produced particle.
The calculation \cite{webpage} uses the same non-perturbative fragmentation
function as used in Ref.~\cite{Cacciari:2012ny},
which is extracted \cite{Cacciari:2005uk}
from $B$-hadron production data in high-energy $e^+e^-$ collisions at LEP
(since the extraction is based on
the resummed FONLL calculation, the direct use of this
non-perturbative fragmentation function in the context of fixed-order calculations, at either NLO or NNLO, is questionable). In the following we comment on the results
of the computation.

In the case of the FONLL calculation of $d\sigma/dp_T$, the non-perturbative fragmentation effects soften the $p_T$ spectrum. They decrease the value of
$d\sigma/dp_T$ in the high-$p_T$ region and, consequently (since the total cross section is unchanged), they increase the value of $d\sigma/dp_T$ in the low-$p_T$
region. The crossover point is at $p_T \sim 4.5$~GeV, where $d\sigma/dp_T$ is almost unchanged. Specifically, we find that the $p_T$ differential cross section decreases by about 25\% at $p_T \sim 50$~GeV, about 20\% at $p_T \sim 20$~GeV and about
13\% at $p_T \sim 10$~GeV. In the low-$p_T$ region, the $p_T$ differential cross section increases by about 10\% at $p_T \sim 2$~GeV.

The non-perturbative fragmentation effects slightly modify the shape of the pseudorapidity cross section (see Fig.~\ref{fig:neweta_av.LHC}).
Including the non-perturbative fragmentation function, we find that the value of
$d\sigma/d\eta$ decreases by about 3\% at $\eta=0$, is almost unchanged in the region close to the peak at $\eta \sim 1.8$ (the position of the peak is slightly shifted forward by $\Delta \eta \sim 0.1$), and increases by about 4\% at $\eta=5$.
Such effects on the shape of $d\sigma/d\eta$ are qualitatively and quantitatively consistent with our expectations. Indeed, as we have discussed in
\ref{app:eta-y} (see Eqs.~(\ref{eq:doublediff})--(\ref{eq:singledip})
and accompanying comments), the shape of $d\sigma/d\eta$ `kinematically' follows
from the shape of $d\sigma/(dp_T dy)$ and is `dynamically' controlled by the size
of the typical value of $p_T/m_T$. The inclusion of the non-perturbative fragmentation function softens the $p_T$ spectrum and it slightly reduces (by few percent) the average value of $p_T$, therefore producing the (few-percent level) effects on
$d\sigma/d\eta$ that we find.

The results of the FONLL calculation can be used to roughly estimate the size of non-perturbative fragmentation effects on fixed-order calculations. We expect effects of ${\cal O}(10\%)$ on $p_T$-dependent distributions (e.g., $d\sigma/dp_T$) in the region from low to intermediate values of $p_T$ (say, $p_T \ltap 20$~GeV), and effects at the few-percent level on $p_T$-inclusive distributions (e.g., $d\sigma/dy$ and
$d\sigma/d\eta$). Comparing these non-perturbative fragmentation effects with perturbative scale uncertainties at NNLO, we see that the fragmentation effects are smaller (although of comparable size) for $p_T$-dependent distributions
(see Figs.~\ref{fig:pT_av.LHC} and \ref{fig:fonll-pt})
and definitely smaller for $p_T$-inclusive distributions
(see Figs.~\ref{fig:y_av.LHC}, \ref{fig:neweta_av.LHC} and \ref{fig:fonll-etay}).
%=============================================================================================

\clearpage
\bibliography{biblio}

\end{document}